\documentclass[aps,prl,twocolumn,superscriptaddress,groupedaddress]{revtex4}
\usepackage{amssymb} \usepackage{color,graphicx} \usepackage{amsmath}
\usepackage{booktabs}
\usepackage{amsbsy} \usepackage{amsthm} 
\usepackage{bm} \usepackage{float} \usepackage{braket}
\usepackage{placeins}
\usepackage[colorlinks=true,citecolor=blue,linkcolor=red,urlcolor=red]{hyperref}
\usepackage[sort&compress]{natbib}
\usepackage{comment}
\usepackage{mathtools}
\usepackage{dcolumn,booktabs}
\newcolumntype{d}[1]{D..{#1}}

\usepackage{array}
\newcolumntype{P}[1]{>{\centering\arraybackslash}p{#1}}
\newcolumntype{M}[1]{>{\centering\arraybackslash}m{#1}}

\newcommand{\bq}{\begin{equation}} \newcommand{\eq}{\end{equation}}
\newcommand{\bqali}{\begin{equation}\begin{aligned}}
\newcommand{\eqali}{\end{aligned}\end{equation}}

\renewcommand\k{{\bf k}}

\newcommand\p{{\bf p}}
\newcommand\q{{\bf q}}
\newcommand\z{{\bf z}}

\newcommand\x{{\bf x}}
\newcommand\y{{\bf y}}

\newcommand\rC{r_\text{\tiny C}}

\newcommand{\g}{g_{\mu\nu}}

\graphicspath{{../Figures/}}

\newcommand{\dd}{\text{d}}
\newcommand{\xtilde}{{\raise.17ex\hbox{$\scriptstyle\sim$}}}

\providecommand{\ket}[1]{\lvert #1 \rangle}
\providecommand{\ave}[1]{\left\langle  #1 \right\rangle}
\providecommand{\bra}[1]{\langle #1 \lvert}

\begin{document}

\author{A. Gundhi}
\email{anirudh.gundhi@phd.units.it}
\affiliation{Department of Physics, University of Trieste, Strada Costiera 11, 34151 Trieste, Italy}
\affiliation{Istituto
Nazionale di Fisica Nucleare, Trieste Section, Via Valerio 2, 34127 Trieste,
Italy}

\author{J.L. Gaona-Reyes}
\email{joseluis.gaonareyes@phd.units.it}
\affiliation{Department of Physics, University of Trieste, Strada Costiera 11, 34151 Trieste, Italy}
\affiliation{Istituto
Nazionale di Fisica Nucleare, Trieste Section, Via Valerio 2, 34127 Trieste,
Italy}

\author{M. Carlesso}
\affiliation{Centre for Theoretical Atomic, Molecular, and Optical Physics,
School of Mathematics and Physics, Queens University, Belfast BT7 1NN, United Kingdom}

\author{A. Bassi}
\affiliation{Department of Physics, University of Trieste, Strada Costiera 11, 34151 Trieste, Italy}
\affiliation{Istituto
Nazionale di Fisica Nucleare, Trieste Section, Via Valerio 2, 34127 Trieste,
Italy}

\title{Impact of dynamical collapse models on inflationary cosmology}

\date{\today}
\begin{abstract}
Inflation solves several cosmological problems at the classical and quantum level, with a strong agreement between the theoretical predictions of well-motivated inflationary models and observations. In this work, we study the corrections induced by dynamical collapse models, which phenomenologically solve the quantum measurement problem, to the power spectrum of the comoving curvature perturbation during inflation and the radiation dominated era. We find that the corrections are strongly negligible for the reference values of the collapse parameters.
\end{abstract}
\maketitle

Quantum Mechanics is very successful, but also problematic. The trouble is that, in its standard formulation, the theory introduces a division between the microscopic quantum world of particles and atoms and the macroscopic  world of classical observers \cite{Weinberg:2017aa}. This division  should not be part of a fundamental theory of nature, even more when the theory is applied to the entire universe, where there are no external observers;  however, removing it proved to be a difficult task, and no shared solution is yet available. 

Models of spontaneous wave function collapse~\cite{Bassi2003,Bassi2013} attempt at removing the arbitrary quantum-classical divide by modifying the Schr\"odinger equation. Suitable nonlinear and stochastic terms are added to the standard quantum dynamics, whose effects scale with the mass of the system. The resulting dynamics is such that microscopic systems behave quantum mechanically, but if they interact to form macroscopic objects, they behave the more classically, the more massive the objects. 

In doing so, collapse models predict a dynamical behavior for matter, which differs from the standard quantum mechanical one: the stochastic terms blur the quantum dynamics, and potentially can be spotted in specific situations. An increasing number of experimental investigations in highly controlled systems~\cite{{Belli:2016aa},{Toros:2017aa},{Vinante:2016aa,Vinante:2017aa},{Piscicchia:2017aa},{Carlesso:2016ac,Carlesso:2018ab},{Helou:2017aa},{Bilardello:2016aa},{Adler:2018aa},{Bahrami:2018aa},Nobakht:2018aa,Carlesso:2018aa,Adler:2019aa,Tilloy:2019aa,{Zheng:2020aa},{Pontin:2020aa},{Vinante:2020aa}} have set significant bounds on the collapse parameters, at the same time leaving much freedom. 

Also cosmology has been used as a playground for collapse models~\cite{Adler2007LU, Adler2007LUC, Adler2009,Martin2012,Lochan2012,Das2014, Leon2015, Mariani2016, Banerjee2017, Martin:2019oqq, Martin2020, Leon2020}. Being the largest and oldest physical system, the Universe can provide relevant information about possible modifications of quantum theory, whose effects would build up during the history of the Universe and would be strongly constrained by the increasing and more and more detailed amount of cosmological observations. {Moreover, there is still an ongoing discussion on how the inflationary quantum fluctuations evolve into classical stochastic variables \cite{Albrecht1994,Polarski1996,Perez2006,Kiefer2009,Sudarsky:2009za,Pinto-Neto2011,Castagnino2014,Okon2015}.}

A recent paper~\cite{Martin2020} claims that a straightforward application of the Continuous Spontaneous Localization (CSL) model \cite{Pearle1989,Ghirardi1990}, the reference collapse model in the literature, to cosmic inflation, {for the most natural choices of the density contrast}, leads to results which are incompatible with experimental evidence.
Specifically, the CSL correction to the power spectrum of the comoving curvature perturbation is calculated, leading to strongly scale-dependant results, which are disproved by observations.  
 To be quantitative, since during inflation the wavelength of the Cosmic Microwave Background (CMB) modes becomes larger than $\rC=10^{-7}$\,m, which is the reference value of one among the  two phenomenological parameters of the CSL model, the authors find  
  $\lambda\ll 5.6\times 10^{-90}$\,s$^{-1}$ as a bound on  the second CSL  parameter. Therefore, the analysis in~\cite{Martin2020} rules out a wide class of CSL theories, since in order for the collapse to be effective, $\lambda > 10^{-20}$\,s$^{-1}$ \cite{Toros:2017aa}.

In this letter, we reconsider the application of CSL to standard cosmology, without entering the debate about its foundations \cite{Perez2006,Sudarsky:2006zx,Sudarsky:2007qz,Sudarsky:2009za,Canate:2013isa, Sudarsky2021}.  We show that a different, yet a very natural choice  of the collapse operator leads to negligible corrections to standard quantum predictions.\\

\noindent\textit{Standard inflationary power spectrum.--}
We briefly overview the standard inflationary dynamics, during which the 
early Universe underwent an accelerated phase of expansion \cite{Guth:1980zm,lyth_liddle_2009,Uzan}, and derive the corresponding power spectrum of the comoving curvature perturbation, for which we will later compute the CSL correction. We refer to \cite{lyth_liddle_2009,Uzan,Riotto} for a more detailed discussion, which is also summarized in the Supplementary Material (SM) \cite{supp}. 

We reconsider the dynamics of the perturbation of the scalar field $\phi$ on a flat Friedmann-Lema\^itre-Robertson-Walker metric $g_{\mu\nu}=a^2(\eta)\eta_{\mu\nu}$ in the presence of a scalar potential $V(\phi)$, where $\eta_{\mu\nu}$ is the Minkowski metric, $a(\eta)$ the scale factor and $\eta$ is the conformal time. In terms of the gauge-invariant Mukhanov-Sasaki variable $\delta\phi_{G}$ \cite{Sasaki1986, Mukhanov1988}, {the action of the scalar perturbations  
is given by \cite{Sasaki1986, Mukhanov1988, MUKHANOV1992203}}
\begin{equation}\label{ActionPertDraft}
S= \frac{1}{2}\int \dd\eta \int\dd \mathbf{x} \left[\dot{u}^2-\delta^{ij}\partial_{i}u\partial_j u+\frac{\ddot{z}}{z}u^2\right],
\end{equation}
{where we have introduced the rescaled field $u(\eta,\x)=a\delta\phi_{\mathrm{G}}$}. Further, $\x$ are the comoving coordinates, $\dot u=\dd u/\dd\eta $, and $z(\eta) = a M_\mathrm{P}\sqrt{2 \epsilon}/c_s$. Here $\epsilon=-H^{-2}\dd H/\dd t$ is the slow-roll parameter, $H=a^{-1}\dd a/\dd t$ is the Hubble parameter, $t$ is the cosmic time and $c_s$ stands for the speed of sound ($c_s=1$ during inflation and $c_s=1/\sqrt{3}$ during the radiation dominated era). During inflation, we will work under the slow-roll approximation assuming $\epsilon\ll 1$ and $\dd\epsilon/ \dd t\approx 0$. {Moreover throughout this letter, we will work in reduced Planck units ($\hbar=1$, $c=1$ and $M^2_{{\mathrm{P}}}= {1}/{8\pi G}$).}

Upon quantization, $\hat{u}(\eta,\x)$ can be expressed in terms of the creation and annihilation operators as
\bq
\hat{u}(\eta,\x)= \int \frac{\dd \k}{(2 \pi)^{3/2}} \left[v_k(\eta)\hat{a}_\k e^{i \k \cdot \x}+ \text{h.c.}\right], \label{uoperator}
\eq
where h.c. denotes the Hermitian conjugate, the creation and annihilation operators satisfy $[{\hat{a}_\k},{\hat{a}_{\k'}^\dagger }]~=~\delta(~\k~-~\k')$, and the modes $v_k(\eta)$ are determined by 
\begin{equation}
v_k(\eta)=\frac{e^{-i k \eta}}{\sqrt{2 k}}\left(1-\frac{i}{k \eta}\right), \label{modeInf}
\end{equation}
under a perfect de Sitter approximation \cite{lyth_liddle_2009, Uzan,Riotto}. 
Given the above definitions, {one can compute quantities of interest such as the variance of the comoving curvature perturbation $\hat{\mathcal R}=\hat u/z$. In the comoving gauge, where the comoving observers measure zero energy flux ($T_{0i}=0$), $\mathcal{R}$  determines the spatial curvature $^{(3)}R_{\mathrm{com}}$ on the hypersurface of constant $\eta$ through $^{(3)}R_{\text{com}}=\frac{4}{a^2}\nabla^2 \mathcal{R}$ \cite{Riotto}.} 

The mean squared quantum expectation value
\bq
\braket{0|\hat{\mathcal R}^2(\x,\eta)|0}=\int \dd\ln k\,\mathcal{P}_{\mathcal R}(k,\eta),\label{R2variance}
\eq
defines the corresponding power spectrum $\mathcal{P}_{\mathcal R}$. The latter reads
\begin{equation}\label{PowSpecComoving}
\mathcal{P}_{\mathcal{R}}(k,\eta)=\frac{c_s^2}{2\epsilon M^2_{\mathrm{P}}}\frac{k^3}{2\pi^2}\frac{|v_k(\eta)|^2}{a^2(\eta)}.
\end{equation}
The modes probed by the CMB  exit the horizon well before the end of inflation. For these modes, the expectation values are indistinguishable
from classical stochastic averages \cite{Albrecht1994, Polarski1996}.
This allows us to use the expression in Eq.~\eqref{R2variance} and relate it to observations \cite{Albrecht1994,Polarski1996,Kiefer2009}.
In general, the power spectrum can be parametrized as $
\mathcal{P}_{\mathcal{R}}=A^*_{\mathcal{R}}({k}/{k_{*}})^{n^*_{\mathcal{R}}-1}$
where the values of $A^*_{\mathcal{R}}$ and $n^*_{\mathcal{R}}$ are determined by Planck data \cite{Akrami2018} at the pivot scale $k_{*}=0.05\,\mathrm{Mpc^{-1}}$ to be
$A_{\mathcal{R}}^{*}= \left(2.099\pm0.014\right)\times 10^{-9}$ and $n_{\mathcal{R}}^{*}={} 0.9649 \pm 0.0042$ at the $68\%$ confidence level. We remark that the expression for $\mathcal{P}_\mathcal{R}$ in Eq.~\eqref{PowSpecComoving} is valid not only for inflation, but also for the radiation dominated era. Both stages will be of interest in this letter.\\

\noindent\textit{Cosmological application of collapse models.--} The application of collapse models to cosmology has been previously considered, with 
motivations ranging from explaining the origin of the cosmic structure \cite{Perez2006, Landau2012, Diez2012, Landau2013, Das2013, Das2014E} and constructing chronogenesis and cosmogenesis models \cite{Pearle2013}, to implementing an effective cosmological constant \cite{Josset2017}. In particular, the {phenomenological parameters of the} CSL model {have been previously} constrained through {a consideration of the heating of the intergalactic medium} \cite{Adler2007LU, Adler2007LUC, Adler2009}, {and spectral distortions of the CMB radiation \cite{Lochan2012}}. Moreover, previous works have studied the modifications due to CSL to the spectra of primordial perturbations at a scalar and tensorial level \cite{Das2014, Leon2015, Mariani2016, Banerjee2017, Martin2020, Leon2020}.

In this work, we study the CSL correction to the power spectrum of the scalar perturbations during inflation and the radiation dominated era. As discussed {in detail} in  SM \cite{supp}, the CSL dynamics can be mimicked by  adding a stochastic Hamiltonian $\hat{H}_{\text{\tiny CSL}}$  to the standard quantum Hamiltonian $\hat{H}_0$. The former  is given by:
\begin{equation}
\hat{H}_{\text{\tiny CSL}}(\eta)=\frac{ \sqrt{\lambda} (4\pi \rC^2)^{3/4}}{m_0}\int \dd \x\,  \xi_\eta(\x) {\hat{{L}}_\text{\tiny CSL}}(\eta,\x), \label{Schrodstoch}
\end{equation}
where $m_0$ is a reference mass set equal to that of a nucleon, {$\hat{{L}}_\text{\tiny CSL}(\eta,\x)$} is the CSL collapse operator yet to be chosen, and
 $\xi_\eta(\x)$ is a white Gaussian noise characterized by zero average $\mathbb{E}[\xi_\eta (\x)]=0$ ($\mathbb E$ denotes the stochastic average over the noise) and correlation function
\begin{equation}
{\mathbb{E}[\xi_\eta (\x)\xi_{\eta'}(\y)] = \frac{\delta(\eta-\eta')}{a(\eta)}\frac{e^{-a^2(\eta)(\x -\y)^2/(4 \rC^2)}}{(4 \pi \rC^2)^{3/2}}}. \label{noiseDC}
\end{equation}
Note that the model is defined in terms of two parameters: $\lambda$ and $\rC$, which are the  collapse rate and the space correlator of the noise respectively. The numerical values of these parameters are constrained by experimental evidence. We will come back on this later.

By considering $\hat H_\text{\tiny CSL}$ as a small perturbation to the full dynamics, one can exploit the standard perturbative approach in the interaction picture and compute the time evolution of a general operator $\hat O(\eta)$ to the leading order. By following standard calculations \cite{Schlosshauer2007}, one can express the expectation value of $\hat O(\eta)$ as  
\begin{equation}
{\overline{O}\equiv \mathbb{E}\braket{\hat{O}(\eta)}}=\braket{\hat{O} (\eta)}_0+ {\delta \overline{O}(\eta)_{\text{\tiny CSL}}},
 \label{expectvalue}
\end{equation}
where we account also for the stochastic average $\mathbb{E}$. Here,  $\braket{\hat{O}(\eta)}_0$ is the expectation value given by standard cosmology, and ${\delta \overline{O}(\eta)_{\text{\tiny CSL}}}$ stands for the CSL correction, which reads
\bqali
{\delta \overline{O}(\eta)_\text{\tiny CSL}} &=-\frac{\lambda}{2 m_0^2}\int_{\eta_0}^\eta \frac{ \dd \eta'}{a(\eta')}\int \dd \x' \dd \x'' e^{-\frac{a^2(\eta'){(\x''-\x')^2}}{4 \rC^2}}\\
\times &{ \bra{\psi} \left[\hat{{L}}
_{\text{\tiny CSL}}^\text{\tiny I}(\eta',\x''),\left[\hat{{L}}_{\text{\tiny CSL}}^\text{\tiny I}(\eta',\x'),\hat{O}^\text{\tiny I}(\eta) \right]\right] \ket{\psi}},
\label{Expect2}
\eqali
where the superscript ``I'' indicates that the operators are evaluated in  the interaction picture and {$\ket{\psi}$ is the} initial state of the system, which we will later set equal to the  {Bunch-Davies vacuum state $\ket{0}$ \cite{Riotto,lyth_liddle_2009,Uzan}}.

We now turn to the specification of the collapse operator $\hat{{L}}_\text{\tiny CSL}(\eta,\x)$. In  standard non-relativistic CSL, $\hat{{L}}_\text{\tiny CSL}(\eta,\x)$ is defined as the mass density operator {$m\hat a^\dag\hat a$} \cite{Pearle1989,Ghirardi1990}. 
 Although to this date there is no satisfactory generalization of collapse models to the relativistic regime \cite{Bassi2003,Bengochea12020, Bengochea22020,Tumulka2006a,Bedingham2011a,Bedingham2011b,Pearle2015,Myrvold2017,Bedingham2019,Tumulka2020,Jones2021a,Jones2021b},  different choices for the collapse operator have been proposed in the cosmological setting. Nevertheless, to our knowledge, all such choices are either linear or, to leading order, linearized in the field perturbation $\hat{u}$ and its conjugate momentum \cite{Martin2020b}. Some authors have chosen the collapse operator to be the rescaled variable $\hat{u}$ itself \cite{Leon2020, Canate:2013isa}, while others have chosen the perturbed matter-energy density $\hat{\delta\rho}$ \cite{Martin2020}, which to leading order, is linear in $\hat{u}$ and $\dot{\hat{u}}$ in standard cosmological perturbation theory. With these choices of the collapse operator, when one describes the cosmological perturbations in the Fourier space, the corresponding modes evolve independently, exactly as in standard cosmology \cite{Uzan, Riotto, MUKHANOV1992203}. However, when generalizing a model, one should retain its characteristic traits. In the case of a generalization of the CSL model, one would like that the collapse operator couples different Fourier modes as in the standard case \cite{Adler:2019aa}, which is not possible when the collapse operator is linear in the fields. 
 
Here, we take the collapse operator to be {$\hat{{L}}_\text{\tiny CSL} (\eta,\x)=\hat{\mathcal{H}}_{0}(\eta,\x)$,} the Hamiltonian density operator of scalar cosmological perturbations,
which is identified by $\hat H_0(\eta)=\int\dd\x\,\hat{\mathcal{H}}_{0}(\eta,\x)$. This choice is a natural, though not unique, relativistic generalization of the non-relativistic mass density. Indeed, in flat spacetime, there is no distinction between the Hamiltonian density of the system and the matter-energy density $\rho$ which, in turn, was considered as a possible generalization of CSL  even in FLRW cosmology \cite{Martin2020}. The role played by gravitational degrees of freedom in the reduction of the quantum mechanical wavefunction is still a subject of active debate \cite{PhysRevA.40.1165,Penrose:1996cv,Di_si_2007,Di_si_2014,Bassi:2017szd}. In this light, and given that the unitary part of the time evolution is governed by the full Hamiltonian of the system, we find it more natural for the collapse operator to be given by $\hat{\mathcal{H}}_{0}(\eta,\x)$. This choice contains contributions from the perturbations of both the standard Einstein-Hilbert term and the matter sector of the full action, while the perturbed matter-energy density $\delta \rho$ is obtained only from the latter \cite{Uzan}.
In addition, even in standard perturbation theory, $\hat{\mathcal{H}}_{0}(\eta,\x)$ is quadratic in the field variable $\hat{u}$ and its conjugate momentum, and therefore is also quadratic in the creation and annihilation operators, in analogy to the mass density of the standard CSL model. \\

\noindent\textit{CSL and inflation.--} During inflation, the Hamiltonian density operator  reads~\cite{Riotto} 
${\hat{\mathcal{H}}_{0}^\text{\tiny I}}(\eta,\x) = \tfrac{1}{2} [\dot{\hat{u}}^2(\eta,\x)+(\nabla \hat{u}(\eta,\x))^2 - \tfrac{2}{\eta^2}\hat{u}^2(\eta,\x)]$,   which is the Hamiltonian density of the scalar perturbations in the Heisenberg picture in standard cosmology, where one does not have additional contributions coming from collapse dynamics. Taking into account  Eq.~\eqref{uoperator}, we have
\bqali
{\hat{\mathcal{H}}_{0}^\text{\tiny I}}(\eta,\x)=   \int\dd \q\dd\p \,&\frac{e^{i(\p + \q)\cdot \x}}{2(2 \pi)^3}  [b_\eta^{\p,\q}\hat{a}_\p \hat{a}_\q + d_\eta^{\p,\q}\hat{a}_{-\q}^\dagger\hat{a}_\p \nonumber\\&+ b_\eta^{\p,\q*}\hat{a}^{\dagger}_{-\p} \hat{a}^{\dagger}_{-\q} +d_\eta^{\p,\q*}\hat{a}^{\dagger}_{-\p} \hat{a}_\q  ],
 \label{HamiltDensityStruc}
\eqali
where we have defined
\bq
\begin{pmatrix}
b_\eta^{\p,\q}\\
d_\eta^{\p,\q}
\end{pmatrix}
=
\begin{pmatrix}
j_\eta^{p,q}\\
l_\eta^{p,q}
\end{pmatrix}-[(\p \cdot \q)+\tfrac{2}{\eta^2}]
\begin{pmatrix}
f_\eta^{p,q}\\
g_\eta^{p,q}
\end{pmatrix},
\eq
and
\begin{equation}
f_{\eta}^{p,q}=v_pv_q,\  g_{\eta}^{p,q}=v_pv_q^*,\ j_{\eta}^{p,q}=\dot{v}_p\dot{v}_q, \  l_{\eta}^{p,q}=\dot{v}_p\dot{v}_q^*. \label{fandg}
\end{equation}
We can now compute the corrections {$\delta \mathcal{P}_{\mathcal{R}}$} to the power spectrum of the curvature perturbation $\hat{\mathcal{R}}$ at the end of  inflation. The first step of the procedure, which is fully reported in  SM \cite{supp}, is to compute the correction to the evolution of $\hat{\mathcal{R}}^2=(\hat u/z)^2$ due to CSL, by evaluating ${\delta\overline{\mathcal{R}^2}(\eta)_\text{\tiny CSL}}$ according to Eq.~\eqref{Expect2}, for the given choice of the collapse operator, {starting from the Bunch-Davies vacuum state} $\ket{0}$. 
We find
\begin{equation}
{\delta \overline{\mathcal R^2}(\eta_e)_{\text{\tiny CSL}}}
=C_{\eta_e}\int \dd \q \dd \p \int_{\eta_0}^{\eta_e}\dd \eta\, e^{-\frac{\rC^2(\p+\q)^2}{4a^2(\eta)}} \mathcal{F}_\eta^{\p,\q},\label{correctionRinf}
\end{equation}
where $\eta_e$ is the conformal time at the end of inflation,
${C_{\eta_e}=-\lambda \rC^3/({8 \epsilon_{\text{inf}} M_{{\mathrm{P}}}^2 a^2(\eta_e) m_0^2 \pi^{9/2}})}$, 
and
\begin{equation}
\mathcal{F}_\eta^{\p,\q}=\mathfrak{Re}\left[b_\eta^{\p,\q}d_\eta^{\q,\p}(f_{\eta_e}^{q,q})^*
-b_\eta^{\p,\q}{(b_{\eta}^{\q,\p})^*}g_{\eta_e}^{p,p}\right]. \label{curlyF}
\end{equation}
To provide an estimate of Eq.~\eqref{correctionRinf}, we first notice that, during inflation, the scale factor is inversely proportional to the conformal time {$a(\eta) \simeq - (H_{\text{inf}} \eta)^{-1}$}, with the Hubble parameter $H_{\text{inf}}$ that can be approximated to a constant. Thus, the {argument of the Gaussian} in Eq.~\eqref{correctionRinf} becomes {$-\rC^2 H_{\text{inf}}^2 {\eta}^2(\p+\q)^2$}. To get a feeling of the orders of magnitudes involved,  the typical  value of $\rC=10^{-7}$\,m $\sim 10^{27}\,M^{-1}_{{\mathrm{P}}}$ is much bigger than $H_{\text{inf}}\sim 10^{-5}\,M_{{\mathrm{P}}}$, so one has that {$\rC H_{\text{inf}}\gg 1$}. This implies that the Gaussian in Eq.~\eqref{correctionRinf} will suppress the integrand for all values of $\p$ and $\q$ except for those where $p\eta\ll 1$ and $q\eta\ll 1$. 
Under such conditions, we can safely expand $\mathcal{F}_\eta^{\p,\q}$ for small $p \eta$ and $q \eta$ and  determine the leading  contribution to Eq.~\eqref{correctionRinf}. {Thus, we} obtain 
\begin{equation}
\mathcal{F}_\eta^{\p,\q}\simeq-\frac{1}{8 p^3 q^4 \eta^8}\left(\frac{4p^3 q \eta^6}{\eta_e^2}+\frac{32 q^4 \eta^6}{9 \eta_e^2} \right). \label{curlyFapproxinf}
\end{equation}
By substituting Eq.~\eqref{curlyFapproxinf} in Eq.~\eqref{correctionRinf}, we find the leading order correction due the CSL  to the mean squared value of the comoving curvature perturbation:
\bq
{\delta \overline{\mathcal{R}^2 }(\eta_e)_\text{\tiny CSL}}=\int \dd\ln k\,{\delta\mathcal{P}_{\mathcal R}}(k,\eta_e),
\eq
with
\begin{equation}
{\delta \mathcal{P}_{{\mathcal{R}}}(k,\eta_e)}\simeq- \frac{17}{36} \frac{\lambda H_{\text{inf}}^3}{\epsilon_{\text{inf}} \pi^2 M_{{\mathrm{P}}}^2  m_0^2}\ln \left( \frac{\eta_e}{\eta_0} \right).  \label{valueInf}
\end{equation}
This is the CSL correction to the power spectrum $\mathcal{P}_\mathcal{{R}}$ computed during inflation. We notice that {$\delta\mathcal{P}_{\mathcal R}(k,\eta_e)$} is independent from $k$ and $\rC$. {This is just an artifact of the leading order expansion in $k\eta$. Indeed, by looking at the exact expression for $\mathcal{F}_\eta^{\p,\q}$ presented in  SM \cite{supp},  it is clear that the exact expression for {$\delta\mathcal{P}_{\mathcal R}(k,\eta_e)$} depends both on $\rC$ and the modes $k$. Moreover,  the leading order expansion was justified by noticing that $\rC H_{\text{inf}}\gg 1$, and therefore indirectly relies on the largeness of $\rC$ compared to the length scale $H_{\text{inf}}^{-1}$.} 

Equation \eqref{valueInf} can be used to set upper bounds on  $\lambda$. To obtain the numerical value of $\delta \mathcal{P}_\mathcal{R}$, we set $\eta_0 \approx -k_*^{-1} \approx -10^{60}M^{-1}_{\mathrm{P}}$, where  $k_{*}=5 \times 10^{-60} M_{\mathrm{P}}$ is the pivot scale, which first crosses the horizon at the efolding number $N_{*}$ satisfying $a(N_*)=k_*/H(N_*)$. In this way the dynamics is restricted up to the time at which the largest scales of interest $2\times 10^{-4}\mathrm{ Mpc}^{-1}\lesssim k_{\mathrm{CMB}}\lesssim 2 \mathrm{Mpc}^{-1}$ exit the horizon during inflation. The e-folding number $N_*$ satisfies $50 \leq N_* \leq 60$ \cite{Akrami2018}. We fix $N_*=60$. The scale factor at the end of inflation $a(\eta_e)$ can then be determined from the relation $a(\eta_e)=a(N_*)\exp (N_*)$.
By setting $\epsilon_{\text{inf}}=0.005$ \cite{Martin2020}, we find
\bq
{\delta \mathcal{P}_{{\mathcal{R}}}}(k,\eta_e)\sim  \lambda/\lambda_\text{\tiny GRW} \times10^{-34},\label{deltaPinflation}
\eq
where $\lambda_\text{\tiny GRW}=10^{-16}\,$s$^{-1}$ \cite{Bassi2003}. {By comparing $\delta \mathcal{P}_\mathcal{R}(k,\eta_e)$ with the observational error of $\mathcal{P}_\mathcal{R}$, which is of order $\approx 10^{-11}$ \cite{Akrami2018}, one obtains an upper bound {$\lambda \lesssim 10^7\,$s$^{-1}$}, which is {17}}
orders of magnitude weaker than the state-of-art result $\lambda\lesssim10^{-10}\,$s$^{-1}$ \cite{Vinante:2020aa}.\\

\noindent\textit{CSL and the radiation dominated era.--}
In standard cosmology, the power spectrum is frozen at the end of inflation for large scales  \cite{Riotto}. However, as pointed out in~\cite{Leon2014,Martin2020}, this may not be the case when the collapse dynamics is also taken into account.
We now calculate the CSL contribution to the evolution of $\hat{\mathcal R}^2$ during the radiation dominated era, which lasts from time $\eta_e$ to $\eta_r=3 \times 10^{60} M_{\mathrm{P}}^{-1}$ {which is estimated by using the fact that $a(\eta_{r})/a(\eta_{e})\approx 3\times 10^{26}$ \cite{RelicCMB}}. {Notice that, as a first approximation, we are not including effects due to the reheating stage \cite{RelicCMB}, and directly consider the radiation-dominated era as following the inflationary one.} 
 
 During this era, the Hamiltonian density reads $
{\hat{\mathcal{H}}_{{0}}^\text{\tiny I}(\eta,\x)=\frac{1}{2}[\dot{\hat{u}}^2 (\eta,\x) + \frac{1}{3}\left(\nabla \hat{u}(\eta,\x) \right)^2 ]}$, where the quantized field $\hat{u}(\eta,\x)$ can still be expressed as in Eq.~\eqref{uoperator}, and related to $\hat{\mathcal{{R}}}$ via $\hat u=z\hat{\mathcal R}$, but now the modes $v_k(\eta)$ are determined as the solutions of the equation
$\ddot{v}_k(\eta)+ \frac{1}{3}k^2 v_k(\eta)=0$.
By solving this equation and matching the curvature perturbation and its derivative with those at the end of inflation \cite{Martin2020}, one can obtain the explicit form for $v_k(\eta)$, which we report in SM \cite{supp}. By following the same choice as for inflation, we fix the collapse operator as {$\hat{{L}}_{\text{\tiny CSL}}=\hat{\mathcal{H}}_{0}$}. {Therefore, in the interaction picture, the collapse operator} can be rewritten as in Eq.~\eqref{HamiltDensityStruc}, where $b_\eta^{\p,\q}$ and $
d_\eta^{\p,\q}$ now follow
\bq
\begin{pmatrix}
b_\eta^{\p,\q}\\
d_\eta^{\p,\q}
\end{pmatrix}
=
\begin{pmatrix}
j_\eta^{p,q}\\
l_\eta^{p,q}
\end{pmatrix}-\frac13(\q \cdot \p)
\begin{pmatrix}
f_\eta^{p,q}\\
g_\eta^{p,q}
\end{pmatrix},
\eq
where $j_\eta^{p,q}$,
$l_\eta^{p,q}$, $f_\eta^{p,q}$ and $
g_\eta^{p,q}$  are defined {in terms of the radiation dominated era mode $v_k(\eta)$ as described in Eq.~\eqref{fandg}.} 
It follows that the CSL correction $\braket{0|\delta\hat{\mathcal R}^2_\text{\tiny CSL} (\eta_r)|0}$ to the mean squared value of the comoving curvature perturbation    generated during the radiation dominated era has the same structure as in Eq.~\eqref{correctionRinf}, with $(\eta_e,\eta_r)$ substituting $(\eta_0,\eta_e)$ and {${C}(\eta_r)=-{\lambda \rC^3}/{(48 M_{{\mathrm{P}}}^2 a^2 (\eta_r) m_0^2 \pi^{9/2}})$ replacing $C(\eta_e)$}.

To quantify the effect, we first notice that during this era the scale factor is proportional to conformal time: $a(\eta)=(\eta - 2 \eta_e)/(H_{\text{inf}} \eta_e^2)$. This expression for the scale factor neglects possible effects during reheating, as it is obtained by matching the well-known expressions for $a(\eta)$ and its derivative during inflation and the radiation dominated era, as it was also derived in \cite{Martin2020}. 
For times $\eta$ close to $\eta_e$, all the modes of cosmological {interest} are outside the horizon and satisfy the condition {$p \eta_e  \ll 1$ and $q\eta_e\ll 1$}. 

{As was the case in the inflationary era, the leading order contribution to $\delta\mathcal{P}_{\mathcal{R}}$ is now obtained by expanding in $p\eta_e\ll 1$, $p\eta\ll 1$ and $p\eta_r\ll 1$. This justification comes from looking at the exact functional form of $\mathcal{F}_\eta^{\p,\q}$ during the radiation dominated era  where the $p\eta$ and $p\eta_r$ terms lead to rapid oscillations of the integrand in the limit $p\eta\gg 1$ and $p\eta_r\gg 1$. This is in contrast to the inflationary era, where the subhorizon contribution $p\eta\gg 1$ is instead suppressed by the exponential term. For more details we refer to the discussion in SM \cite{supp}. Within this approximation, $\mathcal{F}_\eta^{\p,\q}$ is given to leading order by}
\begin{equation}
{\mathcal{F}_\eta^{\p,\q}\simeq - \frac{54}{\epsilon_{\text{inf}}^3 \eta_e^4 q^3}}. \label{curlyFapproxrad}
\end{equation}
By following the procedure delineated above for inflation and reported {in detail in} SM \cite{supp}, we derive the CSL correction to power spectrum of the curvature perturbation at the end of radiation dominated era:
\begin{equation}
{\delta \mathcal{P}_{{\mathcal{R}}}(k,\eta_r)=\frac{9 \lambda H_{\text{inf}}^3 \eta_e^2}{2 \epsilon_{\text{inf}}^3 M_{{\mathrm{P}}}^2 (\eta_r-2  \eta_e)^2 \pi^2 m_0^2}\ln \left( \frac{2\eta_e-\eta_r}{\eta_e} \right)}. \label{valuerad}
\end{equation}
As for the CSL contribution during inflation, we notice that {$\delta \mathcal{P}_{{\mathcal{R}}}(k,\eta_r)$} is independent from $k$ and $\rC$. {This occurs for the same reasons as during the inflationary stage.  In terms of $\lambda_{\text{\tiny GRW}}$, the correction in Eq.~\eqref{valuerad} reads} 
\bq\label{PowRadNumerical}
{\delta \mathcal{P}_{{\mathcal{R}}}(k,\eta_r)}\sim\lambda/\lambda_\text{\tiny GRW} \times10^{-81},
\eq
which can be safely considered as negligible with respect to the contribution obtained during inflation reported in Eq.~\eqref{deltaPinflation}.\\

\noindent\textit{Discussion.--}
Although there is no general consensus on how to generalize the CSL model to a relativistic scenario as required in a cosmological setting, some requirements have already been pointed out \cite{Martin2020, Bengochea12020, Bengochea22020,Leon2020}. We propose a different generalization of the CSL model and study its effects on the  scalar curvature perturbations and corresponding power spectrum. We find that the corrections, when compared to observations \cite{Akrami2018}, provide upper bounds which {by the end of inflation are already} 17 orders of magnitude weaker than those from state-of-art ground {based} experiments \cite{Vinante:2020aa}. {A detailed study concerning possible modifications of other features of the CMB pattern, such as the  presence of acoustic peaks, clearly goes beyond the scope of this letter. However, in the light of the negligible corrections obtained in Eqs.~\eqref{deltaPinflation} and \eqref{PowRadNumerical} to the standard  quantum mechanical power spectrum, we expect our choice of the collapse operator to be fully compatible with observations.} Furthermore, the negligible corrections obtained in our work are in strong contrast to those obtained in Ref.~\cite{Martin2020}. As our calculations show, this difference is a consequence of the fact that the Hamiltonian density of the perturbations is several orders of magnitude smaller than the perturbed matter-energy density in standard cosmology. 
This difference in results for the two choices of the collapse operator is also confirmed in our analysis performed using the perturbative approach within the interaction picture framework \cite{supp}. 

Moreover, we find that the stringent constraints set on the collapse parameters in Ref.~\cite{Martin2020} are not fully self-consistent for the following reason. 
The measure of cosmological perturbations is quantified by the power spectrum which is defined in Eq.~\eqref{PowSpecComoving}. In addition to the standard dynamics, collapse also contributes to the value of the power spectrum with $\delta \mathcal{P}_{{\mathcal{R}}}$ proportional to $\lambda$. Working within  perturbation theory limits the magnitude of the possible collapse induced corrections that can be trusted and hence the range of $\lambda$ that can be observationally constrained. Indeed, if  $\delta \mathcal{P}_{{\mathcal{R}}}$ is much greater than the classical value $\phi^2/M^2_{\mathrm{P}}$ then the assumption of $\delta\phi$ being much smaller than $\phi$,  which is fundamental for the application of linear cosmological perturbation theory, breaks down.  
We suspect that the application of linear perturbation theory in Ref.~\cite{Martin2020} is not valid for the entire range of $\lambda$ values that the authors have excluded. For example, for $\lambda=10^{-16}\,$s$^{-1}$ they find $\delta \mathcal{P}_{{\mathcal{R}}} = 10^{\sim 50}$ that should be compared to the classical value of $\phi^2/M^2_P\sim 1$ which is typically the case during inflation.

 We briefly comment on the claims made by the authors of Ref.~\cite{Martin2020} in their recent work of Ref.~\cite{Martin2021}.  There, it is claimed that the power spectrum vanishes for our choice of the collapse operator. However, as our results show, this is not the case. Moreover, the proof provided in Ref.~\cite{Martin2021} relies on the assumption that the collapse operator leads to a fully localized wavefunction, which does not hold in general, and in fact need not be applied to calculate the variance \cite{supp}. We have considered a physically consistent definition of the power spectrum, and with a well-motivated choice of collapse operator obtained theoretical corrections that are consistent with observations. 

Finally, our work stresses that any eventual validation or discard of the CSL model cannot be made without addressing the issue of its generalization to the relativistic regime, which  -- without question --  is becoming the subject of present and future research.\\

\noindent\textit{Acknowledgments. -- }
AG and JLGR have contributed equally to the work presented in the manuscript.
The authors thank D. Sudarsky, J. Martin and V. Vennin for their useful comments. AG and JLGR thank L. Asprea and C. Jones for useful discussions. AG also thanks L. Ferialdi for several discussions and acknowledges financial support from the University of Trieste and INFN. JLGR acknowledges financial support from The Abdus Salam ICTP. MC and AB acknowledge financial support from the H2020 FET Project TEQ (Grant No. 766900) and the support by grant number (FQXi-RFP-CPW-2002) from the Foundational Questions Institute and Fetzer FranklinFund,  a donor advised fund of Silicon Valley Community Foundation. MC is supported by UK EPSRC (grant nr.~EP/T028106/1).  AB acknowledges financial support from the COST Action QTSpace (CA15220), INFN and the University of Trieste. 


   \onecolumngrid

\appendix

\tableofcontents 

\vspace{1cm}

Our aim is to compute the correction induced by dynamical collapse models to inflationary observables. To determine the standard inflationary observables in the absence of collapse dynamics, the time evolution of a scalar field $\phi$, with a scalar potential $V(\phi)$, is computed in an FLRW universe. 
In addition to the classical dynamics, the perturbation of the scalar field, which couples to the scalar perturbations of the metric $g_{\mu\nu}$, must also be taken into account to explain the observed CMB anisotropies. These perturbations have a quantum mechanical origin, and their quantization procedure resembles closely that of a single particle in a harmonic potential. To reach the anticipated goal, we start with a short recap of the standard discussion on simple harmonic oscillator. Then, we arrive at the action governing the time evolution of the scalar perturbations and construct the inflationary observable of interest. We treat the corrections originating from collapse dynamics as a perturbation to the standard inflationary Hamiltonian of the scalar perturbations, and compute the corrections to the inflationary observable via the interaction picture framework. Finally, we briefly contrast the differences in results which arise when choosing the collapse operator to be linear or quadratic in the perturbations. 

\section{The simple harmonic oscillator}\label{sec:SHO}
The discussion in this section, which partially follows {Ref.~\cite{Baumann}}, gives a short review of the standard discussion on the simple harmonic oscillator (SHO). Of particular interest is the result obtained in Eq.~\eqref{ModeSolution}, which is referred to in the subsequent sections while specifying the quantization procedure of the scalar perturbations of the metric $g_{\mu\nu}$ and the scalar field $\phi$.

The action for the simple harmonic oscillator is given by 
\begin{equation}
S_{\mathrm{\tiny SHO}} = \int \dd t \,\mathcal{L}_{\mathrm{SHO}},\quad\text{where}\quad \mathcal{L}_{\mathrm{SHO}}= \frac{{\dot{x}}^2}{2} - \frac{1}{2}\omega^2 x^2,
\end{equation}
with $x$ satisfying the equation of motion (EOM) 
\begin{equation}\label{SHO}
\ddot{x}+\omega^2 x=0.
\end{equation}
Upon quantization, $x$ is treated as an operator $\hat{x}$, which can be written in terms of the creation $\hat a^\dag$ and annihilation  $\hat a$ operators as
\begin{equation}\label{SHOModeExpansion}
\hat{x} = v(t)\hat{a}+v^{*}(t)\hat{a}^{\dagger}.
\end{equation}
{Equation~\eqref{SHO}} for the position operator $\hat{x}$ implies that the mode $v(t)$ in Eq.~\eqref{SHOModeExpansion} satisfies
\begin{equation}\label{ModeEquation}
\ddot{v}+\omega^2 v=0 {\implies} v(t)= A\exp(-i\omega t)+B\exp(i\omega t).
\end{equation}
Moreover, the commutation relation between $\hat{x}$ and its canonical conjugate $\hat{p}=\frac{\partial\mathcal{L}}{\partial\dot{\hat{x}}}=\dot{\hat{x}}$, yields an additional constraint on the mode $v$
\begin{equation}\label{ConstraintOne}
v{\dot{v}}^{*} - v^{*}\dot{v} = i {\implies} |A|^2 - |B|^2 = \frac{1}{2\omega}.
\end{equation}
Here, we have set $\hbar=1$, and used the fact that $\left[\hat{a},\hat{a}^{\dagger}\right]=1$.
In terms of the creation and annihilation operators, the Hamiltonian $\hat{H}_{\mathrm{SHO}} = \frac{1}{2}\hat{p}^2+\frac{1}{2}\omega^2\hat{x}^2$ reads
\begin{equation}
\hat{H}_{\mathrm{SHO}} = \left(\frac{\dot{v}^2+\omega^2 v^2}{2}\right)\hat{a}\hat{a}+\left(\frac{{\dot{v}}^{*2}+\omega^2v^{*2}}{2}\right)\hat{a}^{\dagger}\hat{a}^{\dagger}+\left(\frac{{|\dot{v}|}^2+\omega^2|v|^2}{2}\right)\left(\hat{a}\hat{a}^{\dagger}+\hat{a}^{\dagger}\hat{a}\right).
\end{equation}
The vacuum state $\ket{0}$ of the annihilation operator $\hat{a}$ is defined as the state that satisfies $\hat{a}\ket{0}=0.$ Demanding that this state 
is an eigenstate of the Hamiltonian yields an additional constraint on the mode $v$
\begin{equation}\label{ConstrainTwo}
\dot{v}^2+\omega^2 v^2=0 {\implies} \dot{v}= \pm i\omega v.
\end{equation} 
The fact that $\dot{v}\propto v$, implies that either $A$ or $B$ in Eq.~\eqref{ModeEquation} is zero.  From Eq.~\eqref{ConstraintOne} it can be seen that  $A$ cannot be zero, and therefore, the solution of the mode $v(t)$ reads
\begin{equation}\label{ModeSolution}
v(t)=\frac{1}{\sqrt{2\omega}}\exp(-i\omega t).
\end{equation}
The standard result obtained in Eq.~\eqref{ModeSolution} specifies the complete time evolution of the operator $\hat{x}(t)$ in terms of the creation and annihilation operators. It will also be useful for completing the quantization of the scalar perturbations in section \ref{sec:PertQuant}.


\section{Single field inflation}\label{sec:Single Field Inflation}

\subsection{Scalar field in FLRW Universe}
Throughout our calculations, we work with a $(-,+,+,+)$ signature, and in reduced Planck units with $\hbar=c=1$, $M_{\mathrm{P}}\equiv 1/\sqrt{8\pi G}$.
The action for an arbitrary homogeneous scalar field $\phi$ with  a generic metric $g_{\mu\nu}$ reads
\begin{equation}
S =\int \dd^4x\sqrt{-g}\left[\frac{M^2_{\mathrm{P}}}{2}R-\frac{1}{2}g^{\mu\nu}\partial_{\mu}\phi\partial_{\nu}\phi - V(\phi)\right], \label{action}
\end{equation}
where the scalar potential $V(\phi)$ is assumed to have the desired properties necessary for slow roll inflationary dynamics. In addition, $R$ and $g$ are the  Ricci scalar and the determinant of the metric tensor $g_{\mu\nu}$, respectively. The variation of the action in Eq.~\eqref{action} with respect to the scalar field $\phi$ leads to the Klein-Gordon equation, given by
\begin{equation}\label{KGGeneral}
\partial_{\mu}\left[\sqrt{-g}g^{\mu\nu}\partial_{\nu}\phi\right] - \sqrt{-g}V,_{\phi}=0.
\end{equation}
As mentioned before, the full inflationary dynamics is studied by decomposing the scalar field and the metric as a classical background part $\bar{\phi}$ and $\bar{g}_{\mu\nu}$, and  their respective perturbations $\delta\phi$ and $\delta g_{\mu\nu}$. Therefore, we have
\begin{equation}\label{background split}
{\phi=\bar{\phi}+\delta\phi\,,\qquad g_{\mu\nu}=\bar{g}_{\mu\nu} + \delta g_{\mu\nu}}.
\end{equation}  
{The classical background dynamics of $\bar{\phi}$ and $\bar{g}_{\mu\nu}$ is computed first, independently of the perturbations $\delta\phi$ and $\delta g_{\mu\nu}$. In order to do so, one has to make a choice for the background metric. Following the standard procedure we take this to be the flat Friedmann-Lema\^itre-Robertson-Walker (FLRW) metric 
\begin{equation}
\g=a^2(\eta)\eta_{\mu\nu},
\end{equation}
where $\eta$ is the conformal time, and $\eta_{\mu\nu}$ the Minkowski metric. On this background, and for a homogeneous background field $\phi(\eta)$, the two Friedmann equations and the Klein-Gordon equation read
\begin{align}
3 M^2_{\mathrm{P}}h^2 ={}& \frac{\dot{\phi}^2}{2}+a^2V\left(\phi\right),\label{Friedmann1}\\
h^2 - \dot{h} ={}& \frac{\dot{\phi}^2}{2 M^2_{\mathrm{P}}},\label{Friedmann2}\\
\ddot{\phi}+2h\dot{\phi}+a^2V,_{\phi}={}&0.\label{KGBack}
\end{align}
In the above equations, we have defined $h\equiv\dot{a}/a$, and the dot denotes a derivative with respect to the conformal time $\eta$. We work with a convention where $\partial_i$ denotes the derivative with respect to the spacial comoving coordinate $x^i$ and $V,_{\phi}\equiv \partial V/\partial\phi$. For a suitable choice of $V(\phi)$, the solutions to Eqs.~\eqref{Friedmann1} - \eqref{KGBack} yield $\phi(\eta)$ and $a(\eta)$ such that, until the end of inflation, $V(\phi(\eta))$ remains (almost) constant and the scale factor $a$ expands exponentially starting from its initial value at the beginning of inflation.

\subsection{Action for the perturbations}
Next, we compute the dynamics of the perturbations of the scalar field $\delta\phi (\eta,\boldsymbol{x})$ and the metric $\delta g_{\mu\nu}(\eta,\boldsymbol{x})$ over the flat FLRW background metric.
To first order, the scalar field perturbation $\delta\phi (\eta,\boldsymbol{x})$ couples only to the scalar perturbations of the metric \cite{Bardeen1980, Uzan}. On a flat FLRW metric, the most general form of the metric including scalar perturbations is given by \cite{Bardeen1980,Sasaki1986,Uzan} 
\begin{equation}\label{metricgeneric}
\mathrm{d}s^2=a^2(\eta)\left[-\left(1+2A\right)\mathrm{d}\eta^2+2 B_{,i}\mathrm{d}x^i\mathrm{d}\eta+\left((1+2\psi)\delta_{ij}+2E,_{ij}\right)\mathrm{d}x^i\mathrm{d}x^{j}\right].
\end{equation}
Two of the functions among $A(\eta,\boldsymbol{x})$, $B(\eta,\boldsymbol{x})$, $\psi(\eta,\boldsymbol{x})$, $E(\eta,\boldsymbol{x})$ can be removed by working in a specific gauge \cite{lyth_liddle_2009,Riotto,Uzan}. In order to derive the EOM  governing the dynamics of perturbations, it is convenient to work in a gauge where $\psi=0$ and $B=0$ \cite{Nakamura:1996da}. In this gauge, the metric reduces to 
\begin{equation}\label{metricGauge}
\mathrm{d}s^2=a^2(\eta)\left[-\left(1+2A\right)\mathrm{d}\eta^2+\left(\delta_{ij}+2E,_{ij}\right)\mathrm{d}x^i\mathrm{d}x^{j}\right].
\end{equation}
After taking the Fourier transform of the perturbations over the comoving coordinates $x^i$, the matrix representation of the metric $g_{\mu \nu}$ reads
\begin{equation}\label{Perturbed Metric}
g_{\mu\nu}(\k)=\bar{g}_{\mu\nu}+\delta g_{\mu\nu}(\k),
\end{equation}
where we have
\begin{equation}\label{Perturbed Metric Matrix}
{\bar{g}}_{\mu\nu}=
\begin{pmatrix}
-a^2(\eta)& 0\\
0               & \delta_{ij} a^2(\eta)
\end{pmatrix},\qquad \delta g_{\mu\nu}(\k)=
\begin{pmatrix}
-2A_{\k} a^2(\eta)& 0\\
0               & \left(-2k_ik_j E_{\k}\right)a^2(\eta)
\end{pmatrix}.
\end{equation} 
In the matrix form the off-diagonal zeros represent the space-time $g_{0i}$ components of the metric.
The perturbed Klein-Gordon equation is obtained by perturbing Eq.~\eqref{KGGeneral}, and retaining only the terms that are first order in the perturbations $A$, $E$, and $\delta\phi$. It is given by
\begin{equation}\label{KGPerturbed1}
\partial_{\mu}\left[(\delta\sqrt{-g})\bar{g}^{\mu\nu}\partial_{\nu}\bar{\phi}\right]+ \partial_{\mu}\left[\sqrt{-\bar{g}}(\delta g^{\mu\nu})\partial_{\nu}\bar{\phi}\right]+\partial_{\mu}\left[\sqrt{-\bar{g}}\bar{g}^{\mu\nu}\partial_{\nu}(\delta\phi)\right]- (\delta\sqrt{-g})V,_{\bar{\phi}}- \sqrt{-\bar{g}}V,_{\bar{\phi}\bar{\phi}}\delta\phi=0,
\end{equation}
where the overhead bar denotes the unperturbed version of the variable. By imposing the homogeneity of the background scalar field $\bar{{\phi}}$ (i.e. $\partial_{i}\bar{\phi}=0$), and by using Eq.~\eqref{Perturbed Metric Matrix} and Eq.~\eqref{KGBack} in Eq.~\eqref{KGPerturbed1} one can simplify the latter, which in Fourier space becomes
\begin{equation}\label{KGPerturbed2}
\ddot{\delta\phi}_{\k}+2h\dot{\delta\phi}_{\k}+(k^2+a^2 V,_{\bar{\phi}{\bar{\phi}}})\delta\phi_{\k}+2A_{\k}a^2V,_{\bar{\phi}} - \dot{\bar{\phi}}(\dot{A_{\k}}+k^2 \dot{E_{\k}})=0\,,
\end{equation}
where $k^2\equiv \k\cdot\k$, and $\delta\phi_{\k}$ is the Fourier component of the inhomogeneous field perturbations $\delta\phi(\x,\eta)$.
Using Einstein's equations $\delta G_{\mu\nu}= \frac{1}{M^2_{\mathrm{P}}}\delta T_{\mu\nu}$ for the scalar perturbations, the metric perturbations $A_{\k}$ and $E_{\k}$ can both be expressed in terms of $\delta\phi_k$ and its derivative as \cite{Nakamura:1996da}
\begin{align}
2A_{\k} ={}&\frac{\dot{\bar{\phi}}\delta\phi_{\k}}{M^2_{\mathrm{P}}h},\label{Einstein1}\\
\dot{A_{\k}}+ k^2\dot{E_{\k}}={}&\frac{\delta\phi_{\k}}{M^2_{\mathrm{P}}}\frac{d}{d\eta}\left(\frac{\dot{\bar{\phi}}}{h}\right)\label{Einstein2}.
\end{align} 
By imposing Eq.~\eqref{Einstein1}, Eq.~\eqref{Einstein2} and Eq.~\eqref{KGBack} in Eq.~\eqref{KGPerturbed2}, we get
\begin{equation}
\ddot{\delta\phi_{\k}}+2h\dot{\delta\phi_{\k}}+\left[k^2+a^2 V,_{\phi\phi}-\frac{1}{M^2_{\mathrm{P}}a^2}\frac{d}{d\eta}\left(\frac{a^2\dot{\phi}^2}{h}\right)\right]\delta\phi_{\k}=0,
\end{equation}
where we have dropped the bar over the unperturbed classical background field $\bar{\phi}$.
Defining the rescaled field variable  $u_{\k}$ as
\begin{equation}
u_{\k}=a \delta \phi_{\k},
\end{equation}
the friction term $2h\dot{u}_{\k}$ disappears, and the EOM for $u_{\k}$ becomes
\begin{equation}\label{MukhSasPart1}
\ddot{u}_{\k}+\left[k^2+a^2V,_{\phi\phi}-\frac{\ddot{a}}{a}-\frac{1}{M^2_{\mathrm{P}}a^2}\frac{d}{d\eta}\left(\frac{a^2\dot{\phi}^2}{h}\right)\right]u_{\k} =0.
\end{equation}
Furthermore, we define the factor $z$ as
\begin{equation}\label{ZFactor}
z \equiv a M_\mathrm{P}\sqrt{2 \epsilon}{/c_s}\,,
\end{equation}
where $c_s$ is the speed of sound (where $c_s=1$ during inflation, and $c_s=1/\sqrt{3}$ during the radiation dominated era). In terms of the cosmic time $dt = a(\eta)d\eta$, the slow roll parameter $\epsilon$ is given by
\begin{equation}\label{SlowRollOne}
\epsilon={} -\frac{1}{H^2}\frac{\dd H}{\dd t},
\end{equation}
where $H$ is the Hubble parameter $H=a^{-1}\frac{d}{dt}a$. Using {Eq.~}\eqref{Friedmann1} and {Eq.~}\eqref{Friedmann2} during inflation we get $z={a\dot{\phi}}/{h}$. Using Eq.~\eqref{KGBack} and Eq.~\eqref{Friedmann2}, as well as the equations resulting from taking their time derivative, Eq.~\eqref{MukhSasPart1} takes the compact form
\begin{equation}\label{MukhSas2}
\ddot{u}_{\k}+\left(k^2 - \frac{\ddot{z}}{z}\right)u_{\k}=0.
\end{equation} 
This equation, governing the dynamics of $u_\k$, can also be obtained from the following action functional
\begin{equation}\label{ActionPert}
\delta S^{(2)} = \frac{1}{2}\int \dd\eta\int \dd \x \left[\dot{u}^2-\delta^{ij}\partial_{i}u\partial_j u+\frac{\ddot{z}}{z}u^2\right]\,,
\end{equation}
which can be computed independently by considering the action of Eq.~\eqref{action} to second order in the scalar perturbations \cite{Mukhanov1988, Sasaki1986, MUKHANOV1992203}. 

We make a short comment here about our choice of the gauge, where $\psi=B=0$. In this gauge, the gauge-invariant field perturbation $\delta\phi_{\text{G}}$, which is defined as \cite{Sasaki1986,Mukhanov1988}
\begin{equation}\label{MukhSasVar}
\delta\phi_{G}\equiv\delta\phi - \dot{\phi}\frac{\psi}{h},
\end{equation}
 is equal to $\delta\phi$ since $\psi=0$. {Thus, in this gauge, one can also identify $u$ with its corresponding gauge-invariant rescaled field $u_G\equiv a\delta\phi_{G}$. In general, one can work with the gauge-invariant variable $\delta\phi_{G}$ defined in Eq.~\eqref{MukhSasVar} from the very beginning and arrive at the gauge-invariant version of the action in Eq.~\eqref{ActionPert} with $u$ replaced with $u_G$ \cite{Sasaki1986, Mukhanov1988, MUKHANOV1992203}}. From now on, we will be working with the gauge-invariant variables, but without explicitly retaining the index `$G$' in the subscript.}

\subsection{Quantizing the perturbations}\label{sec:PertQuant}
In order to allow for sufficient inflation, a successful inflationary model must have an almost flat potential $V(\phi)$, with the scalar field slowly rolling down its small but finite slope \cite{Riotto, lyth_liddle_2009, Uzan}. Equation \eqref{Friedmann1} translates this demand in having a dynamics where the rate of change of the Hubble parameter is rather small. The slow roll dynamics requires $\epsilon$ to be small. The smallness of $\epsilon$ can be qualitatively understood as the condition that the `kinetic energy' of the scalar field $\dot{\varphi}^2/a^2$ is much smaller than the `potential energy' $V(\phi)$. That the kinetic term is small, would follow from {Eq.~\eqref{KGBack}} if one chooses a suitable inflationary potential which has negligible first and second derivatives $\left|M^2_{\mathrm{P}}\left(V,_{\phi}/V\right)^2\right|\ll 1$ and $|M^2_{\mathrm{P}}V,_{\phi\phi}/V|\ll 1$. In conformal time, we have $\epsilon = \dot{\phi}^2/(2M^2_{\mathrm{P}}h^2)$.  Since in the slow-roll approximation both the kinetic term and the acceleration of the scalar field can be neglected, we take $\epsilon\ll 1$ and treat $\epsilon$ as a constant. In this limit one  obtains for a perfect de Sitter spacetime 
\begin{equation}
a(\eta) \approx -1/({H_{\text{inf}}} \eta),
\end{equation}
where we have approximated {$H_{\text{inf}}$} to be a constant, which follows from Eq.~\eqref{Friedmann1} for a flat potential $V$ under {the slow-roll} approximation. Under the same approximation scheme we obtain the standard expressions
\begin{equation}\label{SlowRollApprox}
{\frac{\ddot{z}}{z}\approx \frac{\ddot{a}}{a} \approx \frac{2}{\eta^2}.}
\end{equation}
Using the result of Eq.~\eqref{SlowRollApprox} in Eq.~\eqref{MukhSas2} we get
\begin{equation}\label{MukhSas2PDS}
\ddot{u}_k+\left(k^2 - \frac{2}{\eta^2}\right)u_k =0.
\end{equation}
In analogy to the SHO mode expansion in Eq.~\eqref{SHOModeExpansion}, upon quantization the status of the classical perturbation $u(\eta,\x)$ is raised to that of a field operator $\hat{u}(\eta,\x)$ which can be written as 
\begin{equation}
\hat{u}(\eta,\x)={}\int \frac{\dd\k}{\left(2\pi\right)^{3/2}}\exp(i\k\cdot\x)\hat{u}_{\k}(\eta),\label{CanonicalVariable}
\end{equation}
where the Fourier component $\hat{u}_{\k}(\eta)$ is given in terms of the modes $v_k(\eta)$, and the creation and annilation operators as
\begin{equation}
\hat{u}_{\k}(\eta)={} v_k(\eta)\hat{a}_{\k} + v^*_k(\eta)\hat{a}^\dagger _{-\k}. \label{fieldopu}
\end{equation}
The solution for the modes $v_k(\eta)$ is given by
\begin{equation}
v_k(\eta) = \frac{A(-\frac{\cos(k\eta)}{k\eta}-\sin(k\eta))}{\sqrt{2k}}+\frac{B\left(-\cos(k\eta)+\frac{\sin(k\eta)}{k\eta}\right)}{\sqrt{2k}}.
\end{equation}
 To fix the free parameters $A$ and $B$, one imposes the Bunch-Davies vacuum condition \cite{Riotto,lyth_liddle_2009,Uzan}. Such a condition demands the system to be in the ground state $|0\rangle$ at $\eta\rightarrow -\infty$,  which must then also be an eigenstate of the Hamiltonian corresponding to the action functional for $u$ [cf.~Eq.~\eqref{ActionPert}]. Since the differential equation satisfied by the modes in this limit is identical to that of an SHO [cf. Eq.~\eqref{ModeEquation}], following an analogous procedure as the one described in section \ref{sec:SHO}, we must have
\begin{equation}
\left.v_k(\eta)\right|_{\eta\rightarrow -\infty} = \frac{\exp(-i k\eta)}{\sqrt{2k}}.
\end{equation}
This gives $B=-1$ and $A=i$. In the perfect de Sitter limit, the full solution $v_k(\eta)$ then reads \cite{Uzan,lyth_liddle_2009}
\begin{equation}
v_k(\eta)=\frac{e^{-i\eta k} \left(1-\frac{i}{\eta k}\right)}{(2k)^{1/2}}. \label{modeDS}
\end{equation}

\subsection{Power spectrum}
The power spectrum $\mathcal{P}_{u}$ of the operator $\hat{u}(\eta,\x)$ is defined to be \cite{Riotto,lyth_liddle_2009,Baumann,Uzan}
\begin{equation}
\left\langle0|\hat{u}^2(\eta,\x)\right|0\rangle \equiv \int \dd\ln(k)\,\mathcal{P}_{u}(k,\eta). \label{PowerSpectrumdef}
\end{equation}
The expectation value  $\left\langle0|\hat{u}^2(\eta,\x)\right|0\rangle$ is independent of $\x$. Indeed, if one uses the expression~\eqref{CanonicalVariable} along with Eq.~\eqref{fieldopu} we get
\begin{equation}\label{MeanSqauredExp}
\left\langle0|\hat{u}^2(\eta,\x)\right|0\rangle =\int\int \frac{\dd\k \dd\k'}{(2\pi)^3} e^{i(\k+\k')\cdot \x}v_k(\eta)v^{*}_{k'}(\eta)\langle 0|\hat{a}_{\k}\hat{a}^{\dagger}_{-\k'}|0\rangle.
\end{equation}
Using the commutation relation of the creation and annihilation operators, we see that the expectation value is independent of $\x$ and that the power spectrum $\mathcal{P}_{u}(k,\eta)$ is given by
\begin{equation}
\mathcal{P}_{u}(k,\eta) = \frac{k^3}{2\pi^2}|v_k(\eta)|^2. 
\end{equation}
In the superhorizon limit $k\eta\ll 1$, i.e. when we consider a cosmological perturbation of wavelength larger than the length scale {$1/(aH_\text{inf})$}, the expression for the mode in Eq.~\eqref{modeDS} simplifies to \cite{Uzan}
\begin{equation}
v_k(\eta) \overset{k\eta\ll 1}{\approx} -\frac{i}{\sqrt{2}}\frac{1}{\eta k^{3/2}}.
\end{equation}
Using this result in Eq.~\eqref{PowerSpectrumdef}, we obtain that the power spectrum in the superhorizon limit is given by
\begin{equation}\label{PowerSpectrumModes}
\mathcal{P}_{u}(k,\eta)\approx\frac{1}{\left(2\pi\eta\right)^2}.
\end{equation}
A central quantity of interest for computing inflationary observables is the comoving curvature perturbation $\hat{\mathcal{R}}$, which is related to $\hat{u}$ as
\begin{equation}
\hat{\mathcal{R}}\equiv \hat{u}/z. \label{defcomcurvpert}
\end{equation}
The power spectrum $\mathcal{P}_\mathcal{R}$ of $\hat{\mathcal{R}}$ can therefore be obtained from the power spectrum of $\mathcal{P}_{u}$ as  $\mathcal{P}_{\mathcal{R}}=\mathcal{P}_{u}/z^2$.  More explicitly, it is given by 
\begin{equation}
\mathcal{P}_{\mathcal{R}} \approx \frac{1}{2\epsilon M^2_{\mathrm{P}}}\left(\frac{H}{2\pi}\right)^2.
\end{equation}
We {point out} that, for a perfect de Sitter solution of the modes [cf.~Eq.~\eqref{modeDS}], this result is independent of both the conformal time $\eta$ and the scale $k$. As remarked in the main text, in a more general treatment, where the {slow-roll} dependence is taken into account in the time evolution of the Hubble parameter, the power spectrum acquires a mild scale dependence. The power spectrum is then parametrized as \cite{Riotto,lyth_liddle_2009,Uzan}
\begin{equation}
{\mathcal{P}_{\mathcal{R}}=A^*_{\mathcal{R}}\left(\frac{k}{k_{*}}\right)^{n^*_{\mathcal{R}}-1}.\label{PowerLawPh}}
\end{equation} 
The values of $A^*_{\mathcal{R}}$ and $n^*_{\mathcal{R}}$ are constrained by Planck data \cite{Akrami2018} at the pivot scale $k_{*}=0.05\,\mathrm{Mpc^{-1}}$ to be
\begin{equation}
{A_{\mathcal{R}}^{*}= \left(2.099\pm0.014\right)\times 10^{-9},\quad\text{and}\quad n_{\mathcal{R}}^{*}= 0.9649 \pm 0.0042,\label{PlanckData}}
\end{equation}
at the $68\%$ confidence level.
As mentioned in the main text, we ignore the scale dependence of the power spectrum $\mathcal{P}_{\mathcal{R}}$, and work with a perfect de Sitter solution for the modes reported in Eq.~\eqref{modeDS}. Since the correction induced by collapse dynamics to $A^*_{\mathcal{R}}$ will turn out to be several orders of magnitude below $10^{-9}$, it will make the scale dependence of the correction term irrelevant.

\section{Dynamical collapse models}\label{sec:Dynamical Collapse Model}

Dynamical collapse models are phenomenological models which  modify the standard Schr\"odinger evolution through the addition of nonlinear and stochastic terms. 
The nonlinearity allows the breakdown of superpositions during a measurement, while the stochasticity is necessary in order to avoid faster-than-light signalling  \cite{Bassi2013}. To correctly describe the dynamics of microscopic systems, which are successfully described by quantum mechanics, and that of classical macroscopic systems, one requires that these effects are stronger for larger systems. 
Here, we focus on the Continuous Spontaneous Localization (CSL) model \cite{Pearle1989,Ghirardi1990}, which is the most studied among the dynamical collapse models.
The model is defined through the following stochastic differential equation
\bqali
\dd \ket{\psi}=&\left[-i\hat{H} \dd t + \frac{\sqrt{\gamma}}{m_0} \int \dd \x \left[\hat{M}(\x) - {\braket{\hat{M}(\x)}}\right] \dd W_t(\x) \right.\\
&\left.-\frac{\gamma}{2 m_0^2} \int \dd \x \dd \y {\left[\hat{M}(\x)-\braket{\hat{M}(\x)} \right]G(\x-\y) \left[\hat{M}(\y)-\braket{\hat{M}(\y)}\right]} \dd t \right]\ket{\psi},
 \label{CSL}
\eqali
where $\hat{H}$ is the Hamiltonian of the system,  $\gamma$ is a phenomenological parameter of the model encoding the strength of the collapse process and ${\braket{\,\cdot\,}}$ denotes the expectation value on the state $\ket\psi$. The noise $W_t(\x)$ defined at each point of space is characterized through the correlation
\begin{equation}
\mathbb{E}\left[ \xi_t(\x) \xi_{t'}(\y) \right]=G(\x-\y)\delta(t-t'), \qquad \text{where} \qquad G(\x-\y)=\frac{1}{(4 \pi \rC^2)^{3/2}}e^{-\frac{(\x-\y)^2}{4 \rC^2}}, \label{CSLnoise}
\end{equation}
and $\xi_t(\x)=\dd W_t(\x)/\dd t$. Here, $\mathbb{E}[\cdot]$ indicates the stochastic average, and $\rC$ denotes the second phenomenological parameter of the model. Finally, the operator $\hat{M}(\x)$ in Eq.~\eqref{CSL} is the mass density operator, given by
\begin{equation}
\hat{M}(\x)=
\sum_j m_j \hat{a}_j^\dagger(\x) \hat{a}_j(\x),
\end{equation} 
where the operators $\hat{a}_j^\dagger (\x)$ and $\hat{a}_j(\x)$ are the creation and annihilation operators of a particle of type $j$ in the space point $\x$. The stochastic differential equation for $\ket{\psi}$ in Eq.~\eqref{CSL} leads to the following master equation
\begin{equation}
\frac{\dd \hat{\rho}}{\dd t}=-\frac{\gamma}{2m_0^2} \int \dd \x \dd \y \,G(\x-\y) \left[\hat{M}(\x),\left[\hat{M}(\y),\hat{\rho}\right]\right], \label{CSLmastereq}
\end{equation}
where the density operator $\hat{\rho}$ is defined as $\hat{\rho}=\mathbb{E}[\ket{\psi}\bra{\psi} ]$.  The expectation value $\mathbb{E} [\bra{\psi} \hat{O} \ket{\psi} ]$ of an arbitrary operator $\hat{O}$ can be calculated in terms of the density operator $\hat{\rho}$ in Eq.~\eqref{CSLmastereq} as $\mathbb{E}[\bra{\psi}\hat{O}\ket{\psi} ]=\text{Tr} [\hat{O}\hat{\rho} ]$. Therefore, in order to calculate the expectation values of observables, one may use any unravelling that yields the same master equation of the CSL model. One of these unravellings is  provided by the following linear equation in 
the Stratonovich form 
\begin{equation}
\frac{\dd \ket{\psi}}{\dd t}= -i\left[\hat{H}+ \frac{\sqrt{\gamma}}{m_0}\int \dd \x\, \hat{M}(\x){\xi}_t(\x)\right]\ket{\psi}.  
\end{equation}
Indeed, both the latter equation and Eq.~\eqref{CSL} provide the same master equation for the statistical operator $\hat \rho$ \cite{Bassi2003}.
From the above equation, we can interpret the action of the CSL model as an addition of a stochastic Hamiltonian
\begin{equation}
\hat{H}_{\text{\tiny CSL}}=\frac{\sqrt{\gamma}}{m_0} \int \dd \x \hat{M}(\x) {\xi}_t(\x),
\end{equation}
 to the standard one.

\subsection{Interaction picture framework}
We study the effects of dynamical collapse models by employing a perturbative approach. We split the total Hamiltonian $\hat{H}$ as
\begin{equation}
\hat{H}=\hat{H}_{0}+ \hat{H}_{\text{\tiny CSL}},
\end{equation}
where $\hat{H}_{0}$ is the Hamiltonian of the system and $\hat{H}_{\text{\tiny CSL}}$ is the additional contribution due to dynamical collapse models {in the Schr\"{o}dinger picture}.  
To quantify the correction induced by collapse models, we perform the calculations in the interaction picture. In this picture \cite{Schlosshauer2007}, we identify $\hat{H}_{0}$ as the time dependent background Hamiltonian.  The operators $\hat{O}^\text{\tiny I}$ and the states $\left|\psi^\text{\tiny I}(t)\right\rangle$ are given by
\begin{equation}
\hat{O}^\text{\tiny I} (t)=\hat{U}_{0}^{-1}(t,t_0) \hat{O} \hat{U}_{0}(t,t_0)\,,\qquad {\ket{\psi^\text{\tiny I}(t)}}=\hat{U}_{\text{\tiny CSL}}(t,t_0)\ket{\psi(t_0)}, \label{operint}
\end{equation} 
where 
\begin{equation}
\hat{U}_{0}(t,t_0)=\mathcal{T} \left\{ \exp \left[-i \int_{t_0}^t \dd t' \hat{H}_{0}(t') \right] \right\}\,,\qquad \hat{U}_{\text{\tiny CSL}}(t,t_0)=\mathcal{T} \left\{\exp \left[-i\int_{t_0}^t d t'\hat{H}_{\text{\tiny CSL}}^\text{\tiny I} (t') \right]\right\},
\end{equation}
with $\mathcal{T}$ denoting the time-ordering operator and $\hat{H}_{\text{\tiny CSL}}^\text{\tiny I}$ the Hamiltonian $\hat{H}_{\text{\tiny CSL}}$ in the interaction picture. 
We consider the stochastic Hamiltonian $\hat{H}_{\text{\tiny CSL}}$ to be given by
\begin{equation}
\hat{H}_{\text{\tiny CSL}}(t)=\frac{\sqrt{\gamma}}{m_0} \int \dd \x\, \xi_{t}(\x) \hat{L}_{\text{\tiny CSL}} (t,\x), \label{StochasticHamiltonian}
\end{equation}
where the pair $(t,\x)$ denotes the temporal and spatial coordinates and $\hat{L}_{\text{\tiny CSL}}(t,\x)$ is an operator yet to be specified. The parameter $\gamma$ is the collapse rate of the CSL model and $m_0$ denotes a reference mass taken equal to that of a nucleon. We define $\xi_t(\x)$ as in Eq.~\eqref{CSLnoise}.
Taking into account the time dependence of both the operator $\hat{O}^\text{\tiny I}(t)$ and the state $\ket{\psi(t)}$, and retaining only the leading order term in $\gamma$, we get
\begin{align}
{\braket{\hat{O}}}&=\bra{\psi^\text{\tiny I}(t)}\hat{O}^\text{\tiny I}(t)\ket{\psi^\text{\tiny I}(t)}\\
&\approx \bra{\psi(t_0)}\left[\hat{1}+i\int_{t_0}^t \dd t' \hat{H}_{\text{\tiny CSL}}^\text{\tiny I} (t')  - \int_{t_0}^t \int_{t_0}^{t'}  \dd t' \dd t'' {\hat{H}_{\text{\tiny CSL}}^\text{\tiny I}(t'') \hat{H}_{\text{\tiny CSL}}^\text{\tiny I}(t')} \right]\hat{O}^\text{\tiny I}(t) \left[\hat{1}-i\int_{t_0}^t \dd t' \hat{H}_{\text{\tiny CSL}}^\text{\tiny I} (t') \right.\\\nonumber
&\left.-\int_{t_0}^t \int_{t_0}^{t'} \dd t' \dd t'' \hat{H}_{\text{\tiny CSL}}^\text{\tiny I}(t') \hat{H}_{\text{\tiny CSL}}^\text{\tiny I} (t'')  \right]\ket{\psi(t_0)}\\\nonumber
&=\bra{\psi(t_0)}\left[\hat{O}^\text{\tiny I}(t)-i \int_{t_0}^{t} \dd t' \left[\hat{O}^\text{\tiny I}(t),\hat{H}_{\text{\tiny CSL}}^\text{\tiny I} (t') \right]-\int_{t_0}^t \int_{t_0}^{t'} \dd t' \dd t'' {\left[\hat{H}_{\text{\tiny CSL}}^\text{\tiny I}(t''),\left[\hat{H}_{\text{\tiny CSL}}^\text{\tiny I}(t'),\hat{O}^\text{\tiny I}(t) \right] \right]} \right]\ket{\psi(t_0)}\,.
\end{align}
By taking the stochastic average over all the realizations of the noise, one obtains
\begin{equation}
\begin{split}
&{\overline{O}\equiv\mathbb{E}[\braket{\hat{O}}]}
\approx \bra{\psi(t_0)}\hat{O}^\text{\tiny I}(t)\ket{\psi(t_0)}- \frac{i \sqrt{\gamma}}{m_0} \int_{t_0}^{t} \dd t' \int \dd \x' \mathbb{E}[\xi_t(\x')] \bra{\psi(t_0)}\left[\hat{O}^\text{\tiny I}(t),\hat{L}_{\text{\tiny CSL}}^\text{\tiny I} (t',\x') \right]\ket{\psi(t_0)} \\
&- \frac{\gamma}{m_0^2} \int_{t_0}^{t}\int_{t_0}^{t'} \dd t' \dd t'' \int \dd \x' \int \dd \x''  {\mathbb{E}\left[\xi_{t''}(\x'')\xi_{t'}(\x') \right] \bra{\psi(t_0)}\left[\hat{L}_{\text{\tiny CSL}}^\text{\tiny I}(t'',\x''),\left[\hat{L}_{\text{\tiny CSL}}^\text{\tiny I}(t',\x'),\hat{O}^\text{\tiny I}(t) \right] \right]\ket{\psi(t_0)}}.
\end{split}
\end{equation}
By using Eq~\eqref{CSLnoise}, we get
\begin{equation}\label{StochAvOp}
\begin{split}
{\overline{O}}
& \approx \bra{\psi(t_0)} \hat{O}^\text{\tiny I}(t) \ket{\psi(t_0)}\\
&- \frac{\lambda}{2 m_0^2} \int_{t_0}^{t} \dd t' \int \dd \x' \int \dd \x''  {e^{-\frac{(\x''-\x')^2}{4 \rC^2}} \bra{\psi(t_0)}\left[\hat{L}_{\text{\tiny CSL}}^\text{\tiny I}(t',\x''),\left[\hat{L}_{\text{\tiny CSL}}^\text{\tiny I}(t',\x'),\hat{O}^\text{\tiny I}(t) \right] \right]\ket{\psi(t_0)}},
\end{split}
\end{equation}
where $\lambda=\gamma/(4 \pi \rC^2)^{3/2}$ is the collapse rate. Therefore, we obtain
\begin{equation}
{\overline{O} = \braket{\hat{O}(t)}_0+ \delta \overline{O}(t)_{\text{\tiny CSL}}}, \label{correctexpec}
\end{equation}
where,
\begin{align}\label{PrDeltaPr}
{\braket{\hat{O}(t)}_0}&=\bra{\psi(t_0)}\hat{O}^\text{\tiny I}(t)\ket{\psi(t_0)}, \nonumber\\
{\delta \overline{O}(t)_{\text{\tiny CSL}}}&=- \frac{\lambda}{2m_0^2} \int_{t_0}^{t} \dd t' \int \dd \x' \int \dd \x''  {e^{-\frac{(\x''-\x')^2}{4 \rC^2}} \bra{\psi(t_0)}\left[\hat{\mathcal{H}}_{\text{\tiny CSL}}^\text{\tiny I}(t',\x''),\left[\hat{\mathcal{H}}_{\text{\tiny CSL}}^\text{\tiny I}(t',\x'),\hat{O}^\text{\tiny I}(t) \right] \right]}\ket{\psi(t_0)}.
\end{align}
The term {$\braket{\hat{O}(t)}_0$} represents the expectation value of the operator  $\hat{O}(t)$ in the standard scenario, and the term  ${\delta \overline{O}(t)_{\text{\tiny CSL}}}$ stands for the modification to the expectation value of the operator $\hat{O}$ due to CSL.

\section{Dynamical collapse within a cosmological setting}
  
In what follows, we implement the framework of the previous section within a cosmological setting. Working in conformal time $\eta$ and comoving coordinates $\x$, we set the noise $\xi_\eta(\x)$ to be 
\begin{equation}\label{correlationComoving}
\mathbb{E}[\xi_{\eta}(\x)]=0,\quad\text{and}\quad \mathbb{E}[\xi_{\eta}(\x)\xi_{\eta'}(\y)]=\frac{\delta(\eta-\eta')}{a(\eta')} G(\x-\y), \qquad \text{where} \qquad
G(\x-\y)=\frac{1}{(4 \pi \rC^2)^{3/2}}e^{-\frac{a^{ 2}(\eta)(\x-\y)^2}{4 \rC^2}}.
\end{equation}
Here, the scale factor  $a(\eta)$ is introduced in the above definitions so that $\xi_t(\x_p)$ has the same properties as in the standard CSL model, when expressed in terms of the cosmic time $t$ and the physical coordinates $\x_p$.

For this setting, we have that the expectation value of the operator $\hat O$ can be expressed as in Eq.~\eqref{correctexpec} with  Eq.~\eqref{PrDeltaPr}  now reading
\begin{align}\label{PrDeltaPr2}
{\braket{\hat{O}(\eta)}_0}&=\bra{0}\hat{O}^\text{\tiny I}(\eta)\ket{0}\,, \nonumber\\
{\delta \overline{O}(\eta)_{\text{\tiny CSL}}}&=- \frac{\lambda}{2m_0^2} \int_{\eta_0}^{\eta} \frac{\dd \eta'}{a(\eta')} \int \dd \x' \int \dd \x''  e^{-\frac{a^2(\eta')(\x''-\x')^2}{4 \rC^2}} \bra{0}\left[\hat{\mathcal{H}}_{\text{\tiny CSL}}^\text{\tiny I}(\eta',\x''),\left[\hat{\mathcal{H}}_{\text{\tiny CSL}}^\text{\tiny I}(\eta',\x'),\hat{O}^\text{\tiny I}(\eta) \right] \right]\ket{0}.
\end{align}
In the above expressions, we have set the initial state of the system to be the {Bunch-Davies vacuum state} $\ket{0}$. We will study the modifications due to CSL of the expectation value of the comoving curvature perturbation squared $\hat{\mathcal{R}^2}$ [cf. Eq.~\eqref{defcomcurvpert}] over two cosmological epochs. The first one is the phase of cosmological inflation described in section \ref{sec:inflation} and the second one is that of {the} radiation dominated era described in section \ref{sec:Radiation dominated era}. {These two are separated at $\eta=\eta_{e}$ by a phase of reheating, where $\eta_{e}$ denotes the end of inflation. For a simplified treatment, like in Ref.~\cite{Martin2020}, we assume that the collapse dynamics does not introduce any substantial corrections during this phase connecting the two epochs of interest}. Naturally, in the first epoch $\eta_0$ would correspond to the beginning of inflation, and the correction to ${\overline{\mathcal{R}^2}}$ due to collapse models is computed at the end of inflation. During the radiation dominated epoch, the initial time is taken to be the end of inflation and the correction to ${\overline{{\mathcal{R}}^2}}$ is computed at the end of radiation dominated era $\eta_{r}$.


\subsection{Inflation}
\label{sec:inflation}

We now specify the operator $\hat{{L}}_{\text{\tiny CSL}}$. As motivated in the main text, we take it to be the standard Hamiltonian density of the scalar perturbation $\hat{L}_{\text{\tiny CSL}}=\hat{\mathcal{H}}_{\text{\tiny CSL}}=\hat{\mathcal{H}}_{0}$. Thus, combining Eq.~\eqref{operint} and Eq.~\eqref{StochasticHamiltonian}, we have
\begin{equation}\label{HIDC}
\hat{H}^{\tiny I}_{\text{\tiny CSL}}(\eta)=\frac{\sqrt{\gamma}}{m_0} \int \dd \x \xi_{\eta}(\x) \hat{\mathcal{H}}^{\tiny I}_{\text{\tiny CSL}} (\eta,\x),
\end{equation}
where $\hat{\mathcal{H}}^{\tiny I}_{\text{\tiny CSL}}=\hat{U}_{0}^{-1}\hat{\mathcal{H}}_{0} (\eta,\x)\hat{U}_{0}$ represents the Hamiltonian density of the scalar perturbations in the interaction picture. This coincides with the Hamiltonian density of the scalar perturbations in the Heisenberg picture in standard cosmology, where one does not have additional contributions coming from collapse dynamics. During inflation,  $\hat{\mathcal{H}}^{\tiny I}_{\text{\tiny CSL}}=\hat{\mathcal{H}}^{\text{h}}_{\text{inf}}$, {where the latter} is the standard inflationary Hamiltonian corresponding to the action in Eq.~\eqref{ActionPert}{, and is} given by 
\begin{equation}
\hat{\mathcal{H}}^{\text{h}}_{\text{inf}}=\frac{1}{2} \int \dd \x  \left\{\dot{\hat{u}}^2 (\eta,\x) + \delta^{ij}\partial_{i}{\hat{u}(\eta,\x)}\partial_j {\hat{u}(\eta,\x)}-\frac{2}{\eta^2}\hat{u}^2 (\eta,\x) \right\}\,.
\end{equation}
Imposing the form of the field operator $\hat{u}(\eta,\x)$ {[cf.~Eq.~\eqref{CanonicalVariable}]} we get {the following normal ordered Hamiltonian}
\begin{equation}
\hat{\mathcal{H}}_{\text{\tiny CSL}}^{\tiny I}(\eta,\x)=\int \int \dd \q \dd \p \frac{e^{i(\p + \q) \cdot \x}}{2(2 \pi)^3}\left(b_\eta^{\p,\q}\hat{a}_\p \hat{a}_\q+ d_\eta^{\p,\q}{\hat{a}_{-\q}^\dagger \hat{a}_\p}+ b^{*\p,\q}_\eta \hat{a}_{-\p}^\dagger \hat{a}_{-\q}^\dagger+ d^{*\p,\q}_\eta \hat{a}_{-\p}^\dagger \hat{a}_\q \right),\label{HamiltDensityStruc}
\end{equation}
where we introduced
\begin{equation}
b_\eta^{\p,\q}=j_\eta^{p,q}-\left(\p \cdot \q+\frac{2}{\eta^2}\right)f_\eta^{p,q},\quad d_\eta^{\p,\q}=l_\eta^{p,q}-\left(\p \cdot \q+\frac{2}{\eta^2}\right)g_\eta^{p,q},
\end{equation}
with
\begin{equation}
f_\eta^{p,q}=v_p(\eta)v_q(\eta),\, \quad g_\eta^{p,q}=v_p(\eta) v^*_q(\eta), \quad j_\eta^{p,q}=\dot{v}_p(\eta)\dot{v}_q(\eta), \quad l_\eta^{p,q}=\dot{v}_p(\eta)\dot{v}^*_q(\eta).\label{fandg}
\end{equation}
Now, the quantity we are interested in to be compared with cosmological observations is ${\mathbb{E}[\braket{(\hat{\mathcal{R}}-\braket{\hat{\mathcal{R}}})^2}]}$. However, for a collapse operator which is quadratic in the perturbations, and hence in the creation and annihilation operators, one has that the CSL contribution to $\braket{\hat{\mathcal{R}}}$ is zero, as one can easily deduce from explicit substitution in Eq.~\eqref{PrDeltaPr2}. Thus, one can focus on ${\overline{\mathcal{R}^2}}$ only, which can be obtained from
\begin{equation}\label{Oinflation}
\hat{O}^{\tiny I} (\eta)={\hat{\mathcal{R}}^2(\eta,\x)}=\frac{\hat{u}^2(\eta,\x)}{z^2}=\frac{\hat{u}^2(\eta,\x)}{2 {\epsilon_{\text{inf}}} M_{\mathrm{P}}^2 a^2(\eta)}.
\end{equation}
From Eq.~\eqref{PrDeltaPr2}, the correction induced by the dynamical collapse model is encoded in the term
\begin{equation}
{\delta {\overline{\mathcal{R}^2}(\eta)}_{\text{\tiny CSL}}}=- \frac{\lambda}{2m_0^2} \int_{\eta_0}^{\eta} \frac{\dd \eta'}{a(\eta')}\int \dd \x' \int\dd \x''  {e^{-\frac{a^2(\eta')(\x''-\x')^2}{4 \rC^2}} \bra{0}\left[\hat{\mathcal{H}}_{\text{\tiny CSL}}^{\tiny I}(\eta',\x''),\left[\hat{\mathcal{H}}_{\text{ \tiny CSL}}^{\tiny I}(\eta',\x'),\hat{\mathcal{R}}^2(\eta,\x) \right] \right]}\ket{0}.\label{deltaRinflation}
\end{equation}
{Similarly to the result} in Eq.~\eqref{MeanSqauredExp}, by expressing the Hamiltonian density $\hat{\mathcal{H}}_{\text{\tiny CSL}}^{\tiny I}(\eta,\x)$ and the comoving curvature perturbation $\hat{\mathcal{R}}(\eta,\x)$ in terms of the creation and annihilation operators, explicit calculations show that the correction term ${\delta \overline{\mathcal{R}^2}(\eta)_{\text{\tiny CSL}}}$ is also independent of $\x$. More explicitly, the correction term becomes
\bq\label{deltaRinflationFourier}
{\delta {\overline{\mathcal{R}^2}(\eta)}_{\text{\tiny CSL}}}=-\frac{ \lambda \rC^3}{8 {\epsilon_\text{inf}} M_{\mathrm{P}}^2 m_0^2 a^2(\eta) \pi^{9/2}} \int_{\eta_0}^{\eta}  \frac{\dd \eta'}{a^4(\eta')} \int \int \dd \q \dd \p \,e^{-\frac{\rC^2}{a^2(\eta')}(\q+\p)^2}  \mathfrak{Re} \left[ b_{\eta'}^{\q,\p}\left\lbrace d_{\eta'}^{-\q,-\p}(f_\eta^{q,-q})^*-(b_{\eta'}^{-\q,-\p})^*g_\eta^{q,-q}\right\rbrace \right].
\eq
Notice that, as the exponential is invariant under the interchange of the integration variables $\p$ and $\q$, the properties of the functions $b_{\eta}^{\q,\p}$, $d_{\eta}^{\q,\p}$, $f_{\eta}^{q,p}$, and $g_{\eta}^{q,p}$, allow to write the above result as 
\bq
{\delta \overline{\mathcal{R}^2}(\eta)_\text{\tiny CSL}}=-\frac{ \lambda \rC^3}{8 {\epsilon_{\text{inf}}} M_{\mathrm{P}}^2 m_0^2 a^2(\eta) \pi^{9/2}} \int_{\eta_0}^{\eta}  \frac{\dd \eta'}{a^4(\eta')} \int\int \dd \q \dd \p \,e^{-\frac{\rC^2}{a^2(\eta')}(\q+\p)^2}\mathcal{F}_{\eta'}^{\p,\q},
\label{spectrumcorrInf}
\eq
where
\bq
\mathcal{F}_{\eta'}^{\p,\q}=\mathfrak{Re}\left[b_{\eta'}^{\p,\q}d_{\eta'}^{\q,\p}(f_\eta^{q,q})^*-b_{\eta'}^{\p,\q}(b_{\eta'}^{\q,\p})^*g_\eta^{p,p} \right].\label{curlyFanalytic}
\eq
We are interested in calculating the correction $\delta \mathcal{P}_\mathcal{R}$ at the end of inflation, $\eta=\eta_e$.
By {substituting} $v_k(\eta)$ with the expression in Eq.~\eqref{modeDS}, we obtain the following expression for $\mathcal{F}_{\eta'}^{\p,\q}$
\begin{align}
&\mathcal{F}_{\eta'}^{\p,\q}=\nonumber\\
&\mathfrak{Re}\left[\frac{1}{{8}{\eta'} ^8 p^3 q^4}\left(1+\frac{i}{\text{$\eta $}_e q}\right)^2 e^{-2 i q (\eta' -\text{$\eta $}_e)}\left[\left(\left(-{\eta'} ^2 p^2+i \eta'  p+1\right) \left({\eta'} ^2 q^2-i {\eta'}  q-1\right)-({\eta'}  p-i) ({\eta'}  q-i) \left({\eta'} ^2 (\p \cdot \q)+2\right)\right)\right. \right. \nonumber\\ 
&\left.\left(\left({\eta'} ^2 p^2+i {\eta'}  p-1\right) \left({\eta'} ^2 q^2-i{\eta'}  q-1\right)-({\eta'}  p+i) ({\eta'}  q-i) \left({\eta'} ^2 (\p \cdot \q)+2\right)\right)\right] -\nonumber\\
&\frac{1}{{8}{\eta'} ^8 p^4 q^3}\left(1-\frac{i}{\text{$\eta $}_e p}\right) \left(1+\frac{i}{\text{$\eta $}_e p}\right)\left[ \left(\left(-{\eta'} ^2 p^2+i {\eta'}  p+1\right) \left({\eta'} ^2 q^2-i {\eta'}  q-1\right)-({\eta'}  p-i) ({\eta'}  q-i) \left({\eta'} ^2 (\p \cdot \q)+2\right)\right)\right.\nonumber\\
&\left. \left.\left(-\left({\eta'} ^2 p^2+i {\eta'}  p-1\right) \left({\eta'} ^2 q^2+i {\eta'}  q-1\right)-({\eta'}  p+i) ({\eta'}  q+i) \left({\eta'} ^2 (\p \cdot \q)+2\right)\right)\right]\right]. \label{curlyFanalytInf}
\end{align}
For times close to the end of inflation, the condition $q \eta' \ll 1$ is satisfied for the modes cosmological interest \cite{Uzan}. At earlier times, if this condition is not satisfied, then the exponential appearing in Eq.~\eqref{spectrumcorrInf} suppresses the corresponding contributions. Indeed, during inflation the  exponential function in Eq.~\eqref{spectrumcorrInf} becomes $\exp{\left(-\rC^2 {H_{\text{inf}}}^2 {\eta^{\prime}}^2(\p+\q)^2\right)}$. To have an estimate of the orders of magnitudes involved, we notice that since the GRW value of $\rC\sim 10^{27}M^{-1}_{\mathrm{P}}$ \cite{Ghirardi1986} is much bigger than ${H_{\text{inf}}}\sim 10^{-5}M_{\mathrm{P}}$, it is safe to assume that $\rC {H_{\text{inf}}}\gg 1$ during inflation. Therefore, we can safely expand Eq.~\eqref{curlyFanalytInf} in powers of $q \eta'$ (or $p \eta'$), which to leading order gives
\begin{equation}
\mathcal{F}_{\eta'}^{\p,\q}=\frac{1}{8 p^3q^4{\eta'}^8}\left(-\frac{2q^4{\eta}^4_e}{9}+\frac{16 q^4 \eta_e \eta'^3}{9} -\frac{4p^3q{\eta'}^6}{{\eta}^2_e}-\frac{32q^4{\eta'}^6}{9\eta_e^2}\right). \label{curlyFapproxInf}
\end{equation}
Here, the leading order expression presented above contains only the terms that would survive after computing the integral in Eq.~\eqref{spectrumcorrInf}. That is, terms which are symmetrical in $\p$ and $\q$ but appear with opposite signs would not contribute to the integral and therefore do not appear in the effectively leading order expression {in Eq.~}\eqref{curlyFapproxInf}. Since $\eta_{e}\ll\eta'$,  the last two terms on the RHS in Eq.~\eqref{curlyFapproxInf} give the dominant contribution to the corrections with $\mathcal{F}_{\eta'}^{\p,\q}\approx-1/(2q^3\eta^2_e\eta'^2) - 4/(9\eta^2_e\eta'^2p^3)$. Since $\p$ and $\q$ are dummy integration variables, to leading order $\mathcal{F}_{\eta'}^{\p,\q}$ effectively depends only on $q$ (or $p$). After completing the {$\p$} integral  in Eq.~\eqref{spectrumcorrInf}, which now becomes a standard three dimensional Gaussian integral,  we get
\begin{equation}
{\delta\overline{\mathcal{R}^2}(\eta_e)_{\text{\tiny CSL}}}\approx-\frac{17}{36}\frac{\lambda {H_{\text{inf}}}^3}{\pi^2 {
\epsilon_{\text{inf}}} M_{\mathrm{P}}^2 m_0^2}\int_{\eta_0}^{\eta_e} { \dd}\ln{\eta} \int {\dd} \ln{q}\,.
\end{equation}
Following the definition of the power spectrum $\mathcal{P}$ [cf. Eq.~\eqref{PowerSpectrumdef}], we identify the correction $\delta\mathcal{P}_{\mathcal{R}}$ to the power spectrum $\mathcal{P}_\mathcal{R}$ with ${\delta\overline{\mathcal{R}^2}(\eta_e)_{\text{\tiny CSL}}}=\int \dd \ln q \,\delta\mathcal{P}_\mathcal{R}$. Therefore we obtain   
\begin{equation}
\delta \mathcal{P}_\mathcal{R}\approx-\frac{17}{36} \frac{\lambda {H_{\text{inf}}}^3}{\pi^2 {\epsilon_{\text{inf}}} M_{\mathrm{P}}^2 m_0^2}\ln \left( \frac{\eta_e}{\eta_0}\right).
\end{equation}

\subsection{Radiation dominated era}\label{sec:Radiation dominated era}

During the radiation dominated era, instead of Eq.~\eqref{MukhSas2}, we have
\begin{equation}\label{EOMRad}
\ddot{u}_k+\frac{1}{3}k^2 u_k=0,
\end{equation}
as in general, the action for the rescaled variable $u$ reads \cite{Mukhanov2005}
\begin{equation}
\delta S^{(2)}=\frac{1}{2}\int \dd \eta \int \dd \x  \left[\dot{u}^2-c_s^2\delta^{ij}\partial_i u\partial_j u + \frac{\ddot{z}}{z}u^2 \right], \label{Actiongeneral}
\end{equation}
with $c_s$ being the speed of sound defined before.
{As stated before, for simplicity we neglect in our analysis the reheating stage \cite{Martin2020}}. The scale factor during the radiation dominated era can then be approximated as
\begin{equation}
a(\eta)=\frac{1}{{H_\text{inf}} \eta_e^2}(\eta-2 \eta_e)\,. \label{scalefactorrad}
\end{equation}
By using this expression of $a(\eta)$ in the definition of $\epsilon$, one can show that $\epsilon=2$ during the radiation dominated era. 
{From this result, as well as the linear dependence of the scale factor on $\eta$, the definition of $z$ in Eq.~\eqref{ZFactor} leads to $\ddot{z}=0$.} This explains the absence of a term proportional to $\ddot{z}$ in the EOM~\eqref{EOMRad}. 
We can decompose $u$ in terms of the modes $v_k(\eta)$ as in Eq.~\eqref{fieldopu}, where now the mode satisfies 
\begin{equation}
\ddot{v}_k(\eta)+\frac{1}{3}k^2 v_k(\eta)=0.
\end{equation}
Following the approach of Ref.~\cite{Martin2020}, the initial conditions required to specify the solution for the modes during the radiation dominated era, are fixed by matching the curvature perturbation and its derivative at the end of inflation.   The full solution  of $v_k(\eta)$ then becomes \cite{comment}
\begin{equation}
v_k(\eta)\!\!=\!\!\frac{\sqrt{3}}{2 \eta_e^2 \sqrt{{\epsilon_{\text{inf}}}} k^{5/2}}e^{-ik \eta_e}\!\left\{\left[(1\!+\!\sqrt{3})(k \eta_e)^2\!-\! \sqrt{3}\!-\!i(1\!+\!\sqrt{3})k \eta_e \right]e^{-ik \frac{\eta-\eta_e}{\sqrt{3}}}\!\!+\!\!\left[(1\!-\!\sqrt{3})(k\eta_e)^2 \!+\! \sqrt{3}\!-\!i(1\!-\!\sqrt{3}) k \eta_e\right]e^{ik \frac{\eta-\eta_e}{\sqrt{3}}} \right\}. \label{radiationmode}
\end{equation} 
Now, one obtains the Hamiltonian density during the radiation dominated era
from Eq.~\eqref{Actiongeneral}. This reads  
\begin{equation}
\hat{\mathcal{H}}_{\text{\tiny CSL}}^{\tiny I}(\eta,\x)=\hat{\mathcal{H}}_{\text{rad}}^\text{h}(\eta,\x)=\frac{1}{2}\left(\dot{\hat{u}}^2 (\eta,\x) + \frac{1}{3}\delta^{ij}\partial_i\hat{u}(\eta,\x)\partial_j\hat{u}(\eta,\x) \right). \label{HamDensRad}
\end{equation}
From the expression of the operator $\hat{u}(\eta,\x)$ in Eq.~\eqref{CanonicalVariable}, and its decomposition in terms of the modes $v_k(\eta)$ of Eq.~\eqref{fieldopu}, straightforward calculations lead to an expression for $\hat{\mathcal{H}}_{\text{\tiny CSL}}^{\tiny I}(\eta,\x)$ that has the structure of Eq.~\eqref{HamiltDensityStruc}, but with the functions $b_\eta^{\q,\p}$ and $d_\eta^{\q,\p}$ modified as
\begin{equation}
b_\eta^{\q,\p}=j_\eta^{q,p}-\frac{1}{3}(\q \cdot \p)f_\eta^{q,p},\quad d_\eta^{\q,\p}=l_\eta^{q,p}-\frac{1}{3}(\q \cdot \p)g_\eta^{q,p}, \label{Frad}
\end{equation}
with the functions appearing on the RHS of Eq.~\eqref{Frad} defined as in Eq.~\eqref{fandg}. 
During the radiation dominated era, the same operator $\hat{O}^{\tiny I}(\eta)={\hat{\mathcal{R}}^2}(\eta,\x)$ defined in Eq.~\eqref{Oinflation} reads 
\begin{equation}
{\hat{\mathcal{R}}^2}(\eta,\x)=\frac{\hat{u}^2 (\eta,\x)}{{12} M_P^2 a^2(\eta)}, \label{perturboprad}
\end{equation}
where the scale factor is given by Eq.~\eqref{scalefactorrad}. We have used the fact that $\epsilon(\eta_{e}\leq\eta\leq \eta_{r})=2$ and $c^2_s=1/3$ during the radiation dominated era (with $\eta_r$ denoting the conformal time at the end of this stage). From Eq.~\eqref{PrDeltaPr2}, the contribution to the modification of the comoving curvature power spectrum during the radiation dominated era is given by Eq.~\eqref{deltaRinflation}, where one substitutes $\eta_0$ with $\eta_e$ and $\eta$ with $\eta_r$.

Using Eqs.~\eqref{HamDensRad},~\eqref{perturboprad} and~\eqref{scalefactorrad}, and calculating the double commutator explicitly, we obtain
\begin{equation}
{\delta \overline{\mathcal{R}^2}(\eta_r)_\text{\tiny CSL}}=-\frac{ \lambda \rC^3}{{48} M_{\mathrm{P}}^2 m_0^2 a^2(\eta_r) \pi^{9/2}} \int_{\eta_e}^{\eta_r}  \frac{\dd \eta'}{a^4(\eta')} \int\int \dd \q \dd \p\, e^{-\frac{\rC^2}{a^2(\eta')}(\q+\p)^2}\mathcal{F}_{\eta'}^{\p,\q},
\label{spectrumcorrRad}
\end{equation}
where, using Eqs.~\eqref{fandg} and~\eqref{Frad}, the function $\mathcal F_{\eta'}^{\p,\q}$ [cf. Eq~\eqref{curlyFanalytic}] is given by 
\begin{align}\label{CurlyFRad}
\begin{split}
&\mathcal{F}_{\eta'}^{\p,\q}=-\frac{1}{8 p^5 q^5 \epsilon ^3 \eta_e^4}9 e^{-\frac{2 i (p (\eta' -\eta_e)+q (\eta' +\eta_r-2 \eta_e))}{\sqrt{3}}} \left( 4 e^{\frac{2 i (p+q) (\eta' -\eta_e)}{\sqrt{3}}} (\p \cdot \q) \left(-2 q \eta_e \left(q^3 \eta_e^3-2 q \eta_e+\sqrt{3} i\right)-3\right) p^5 \right.\\
&\left.+4 e^{\frac{2 i (p (\eta' -\eta_e)+q (\eta' +2 \eta_r-3 \eta_e))}{\sqrt{3}}} (\p \cdot \q) \left(2 q \eta_e \left(-q^3 \eta_e^3+2 q \eta_e+\sqrt{3} i\right)-3\right) p^5\right.\\
&\left.+2 e^{\frac{2 i (p+2 q) (\eta' -\eta_e)}{\sqrt{3}}} q^3 \left(p^2 q^2-(\p \cdot \q)^2\right) \left(4 p^4 \eta_e^4-2 p^2 \eta_e^2+3\right)+2 e^{\frac{2 i (p \eta' +2 q \eta_r-(p+2 q) \eta_e)}{\sqrt{3}}} q^3 \left(p^2 q^2-(\p \cdot \q)^2\right) \left(4 p^4 \eta_e^4-2 p^2 \eta_e^2+3\right)\right.\\
&\left.+4 e^{\frac{2 i (p (\eta' -\eta_e)+q (\eta' +\eta_r-2 {\eta_e}))}{\sqrt{3}}} \left(4 p^4 q^3 (p q+\p \cdot \q)^2 \eta_e^4-2 p^2 q^2 \left(2 (\p \cdot \q) p^3+q^3 p^2+q (\p \cdot \q)^2\right) \eta_e^2\right.\right.\\
&\left.\left.+3 \left(2 (\p \cdot \q) p^5+q^5 p^2+q^3 (\p \cdot \q)^2\right)\right)+e^{\frac{4 i (p \eta' +q \eta_r-(p+q) \eta_e)}{\sqrt{3}}} q^3 (-p q+\p \cdot \q)^2 \left(2 p \eta_e \left(p^3 \eta_e^3-2 p \eta_e-\sqrt{3}i\right)+3\right) \right.\\
&\left.+e^{\frac{4 i (p+q) (\eta' -\eta_e)}{\sqrt{3}}} q^3 (p q+\p \cdot \q)^2 \left(2 p \eta_e \left(p^3 \eta_e^3-2 p \eta_e-\sqrt{3}i \right)+3\right) \right.\\
&\left.+2 e^{\frac{2 i (2 p \eta' +q \eta' +q \eta_r-2 (p+q) \eta_e)}{\sqrt{3}}} q^3 \left(p^2 q^2-(\p \cdot \q)^2\right) \left(2 p \eta_e \left(p^3 \eta_e^3-2 p \eta_e-\sqrt{3} i\right)+3\right)\right.\\
&\left.+e^{\frac{4 i q (\eta' -\eta_e)}{\sqrt{3}}} q^3 (-p q+\p \cdot \q)^2 \left(2 p \eta_e \left(p^3 \eta_e^3-2 p \eta_e+\sqrt{3} i\right)+3\right)+e^{\frac{4 i q (\eta_r-\eta_e)}{\sqrt{3}}} q^3 (p q+\p \cdot \q)^2 \left(2 p \eta_e \left(p^3 \eta_e^3-2 p \eta_e+\sqrt{3} i\right)+3\right)\right.\\
&\left.+2 e^{\frac{2 i q (\eta' +\eta_r-2 \eta_e)}{\sqrt{3}}} q^3 \left(p^2 q^2-(\p \cdot \q)^2\right) \left(2 p \eta_e \left(p^3 \eta_e^3-2 p \eta_e+\sqrt{3} i\right)+3\right)\right).
\end{split}
\end{align}
We now consider, as in the inflationary era, the expansion of $\mathcal{F}_{\eta'}^{\p,\q}$ in powers of $q \eta'$, $q\eta_e$ and $q\eta_r$. The leading order term reads 
\begin{equation}\label{CurlyFRadApprox}
\mathcal{F}_{\eta'}^{\p,\q}\approx-{\frac{54}{q^3 {\epsilon_{\text{inf}}^3} \eta_e^4}}\,,
\end{equation}
where the terms that are symmetric in $\p$ and $\q$ but appear with opposite signs have been discarded in the effectively leading order expression {of Eq.}~\eqref{CurlyFRadApprox} as they would yield {a} zero contribution to the integral in Eq.~\eqref{spectrumcorrRad}. Using the leading order expansion {of Eq.}~\eqref{CurlyFRadApprox} in Eq.~\eqref{spectrumcorrRad}, as was the case for the inflationary era, the $\p$ integral becomes a standard three dimensional Gaussian integral. After completing the $\p$ integral first we get 
\begin{equation}
{\delta \overline{\mathcal{R}^2}(\eta_r)_{\text{\tiny CSL}}}\approx\frac{9 \lambda {H_\text{inf}}^3 \eta_e^2}{{2} M_{{\mathrm{P}}}^2  {\epsilon_\text{inf}^3} (\eta_r-2 \eta_e)^2  \pi^2 m_0^2} \int_{\eta_e}^{\eta_r}\dd \ln(\eta'-2 \eta_e)\int \dd \ln q.
\end{equation}
Therefore, the correction $\delta \mathcal{P}_\mathcal{R}$ to the power spectrum $\mathcal{P}_\mathcal{R}$ is given by
\begin{equation}
\delta \mathcal{P}_\mathcal{R}\approx\frac{9 \lambda {H_\text{inf}^3} \eta_e^2}{2 M_{{\mathrm{P}}}^2 {\epsilon_{\text{inf}}^3} (\eta_r-2 \eta_e)^2  \pi^2 m_0^2}\ln \left(\frac{2 \eta_e-\eta_r}{\eta_e} \right).
\end{equation}
We must note that strictly speaking, due to the coupling between the modes in Eq.~\eqref{spectrumcorrRad}, the cosmological modes which might be outside the horizon in the outer $\p$ integral can also receive contributions from the subhorizon modes which satisfy $p\eta'\gg 1$, for which the approximation scheme breaks down. This was not a problem  during inflation because in that era the scale factor is given by $a=-\frac{1}{{H_{\text{inf}}}\eta'}$. Consequently the exponential function in Eq.~\eqref{spectrumcorrInf} becomes $\exp{\left(-\rC^2 {H_{\text{inf}}}^2 {\eta^{\prime}}^2(\p+\q)^2\right)}$, and as explained before, since typically $\rC\gg 1/{H_{\text{inf}}}$, it becomes necessary to have $p^2{\eta'}^2\ll 1$ (and $ q^2{\eta'}^2\ll 1$) for the integrand to be non-zero. Due to the modified functional dependence of the scale factor during the radiation dominated era given in Eq.~\eqref{scalefactorrad}, the exponential function becomes $\exp{\left(\left(\p+\q\right)^2\eta^2_{e} {H_\text{inf}}^2 \rC^2\frac{\eta^2_{e}}{(\eta'-2 \eta_{e})^2}\right)}$. Clearly, it is no longer necessary to have $p\eta'\ll 1$ in order for the exponent to be non-zero. While the condition $p\eta_{e}\ll 1$ still remains valid, as all the modes of interest are outside the horizon at the end of inflation, the expansion in $p\eta'$ and $p\eta_{r}$ needs further justification. In order to see this we notice that the exact expression in Eq.~\eqref{CurlyFRad} depends on $\eta'$ and $\eta_r$ only via the terms $p\eta'$ and $p\eta_r$ appearing in the oscillating phases. Thus, when a mode enters the horizon during the radiation dominated era (i.e. the mode $p$ now satisfies $p\eta'\gg 1$ compared to $p\eta_e\ll 1$ at the end of inflation), this phase is expected to oscillate strongly and would not yield any significant contribution to the integrand. Moreover, the $a^4$ factor in the denominator would also suppress the contribution for a given mode $p$ at a later time, when $p$ enters the horizon and $p\eta'\gg 1$. Therefore, the assumption that $p\eta'\ll 1$ and $p\eta_{r}\ll 1$
for all modes $p$ and at all times $\eta'$ is expected to provide an upper bound on the integral. 

\section{Linearized collapse operator}

Let us consider a generic collapse operator which is linear in the perturbation $\hat{u}(\eta,\x)$ and hence in the creation and annihilation operators. For such a collapse operator  $\hat{l}(\eta,\x)$ one can write
\begin{equation}\label{LinColOp}
\hat{l}(\eta,\x)=\int \frac{\dd \k}{(2\pi)^{3/2}} e^{i \k \cdot \x} \left(\chi_{\k}(\eta) \hat{a}_\k + \chi^*_{-\k} (\eta) \hat{a}_{-\k}^\dagger \right),
\end{equation}
where $\chi_\k (\eta)$ is a suitable function.

We calculate the correction to the expectation value of the comoving curvature perturbation $\hat{\mathcal{R}}(\eta,\x)$ due to the linearized collapse operator $\hat{l}(\eta,\x)$.
In terms of the creation and annihilation operators, $\hat{\mathcal{R}}(\eta,\x)$ is given by (after normal ordering)
\begin{equation}
\hat{\mathcal{R}}^2(\eta,\x)=\frac{\hat{u}^2(\eta,\x)}{z^2}=\frac{1}{2 {\epsilon_{\text{inf}}} M_{\mathrm{P}}^2 a^2(\eta)}\int \frac{\dd \p \dd \q}{(2\pi)^3} e^{i (\p+\q) \cdot \x}\left(f_{\eta}^{\p,\q}\hat{a}_\p \hat{a}_\q + g_{\eta}^{\p,\q}\hat{a}_{-\q}^\dagger \hat{a}_\p + g^{*\p,\q}_{\eta}\hat{a}_{-\p}^\dagger \hat{a}_\q + f^{*\p,\q}_{\eta}\hat{a}_{-\p}^\dagger \hat{a}_{-\q}^\dagger \right),
\end{equation}
where $f,g$ are defined in Eq. \eqref{fandg}.

Following the prescription described before [cf. Eq.~\eqref{PrDeltaPr2}], the correction due to the collapse dynamics is given by
\begin{equation}
{\delta \overline{\mathcal{R}^2}(\eta,\z)}= - \frac{\lambda}{4 m_0^2 \epsilon_{\text{inf}} M_{\mathrm{P}}^2 a^2(\eta)} \int_{\eta_0}^\eta \frac{\dd \eta'}{a(\eta')} \int  \dd \x \dd \y e^{-\frac{a^2(\eta')}{4 \rC^2}(\x-\y)^2} \bra{0} \left[\hat{l}(\eta',\x),\left[\hat{l}(\eta',\y),\hat{u}^2(\z,\eta) \right]\right] \ket{0}.
\end{equation}
Therefore, we need to calculate the double-commutator
\begin{equation}
\begin{split}
&\left[\hat{l}(\eta',\x),\left[\hat{l}(\eta',\y),\hat{u}^2 (\eta,\z)\right]\right]= \int \frac{\dd \k_1 \dd \k_2 \dd \q  \dd \p}{(2\pi)^6} e^{i \p \cdot \x}e^{i \q \cdot \y}e^{i (\k_1+\k_2)\cdot \mathbf{z}} \\
&\left[ \chi_\q \hat{a}_\q + \chi^*_{-\q} \hat{a}_{-\q}^\dagger, \left[\chi_\p \hat{a}_\p + \chi^*_{-\p} \hat{a}_{-\p}^\dagger, f^{\k_1, \k_2}\hat{a}_{\k_1}\hat{a}_{\k_2}+g^{\k_1\k_2}\hat{a}_{-\k_2}^\dagger \hat{a}_{\k_1}+g^{*\k_1, \k_2}\hat{a}_{-\k_1}^\dagger \hat{a}_{\k_2}+ f^{*\k_1, \k_2}\hat{a}_{-\k_1}^\dagger \hat{a}_{-\k_2}^\dagger \right]\right].
\end{split}\end{equation}
Explicitly, for the first term we obtain
\begin{equation}
\left[\chi_\q \hat{a}_\q + \chi^*_{-\q} \hat{a}_{-\q}^\dagger, \left[\chi_\p \hat{a}_\p + \chi^*_{-\p} \hat{a}_{-\p}^\dagger, f^{\k_1, \k_2}\hat{a}_{\k_1}\hat{a}_{\k_2}\right]\right]=\chi^*_{-\q} \chi^*_{-\p} f^{\k_1,\k_2} \left(\delta(\k_2+\p)\delta(\k_1+\q)+\delta(\k_1+\p)\delta(\k_2+\q) \right),
\end{equation}
for the second term
\begin{equation}
\begin{split}
&\left[\chi_\q \hat{a}_\q + \chi^*_{-\q} \hat{a}_{-\q}^\dagger, \left[\chi_\p \hat{a}_\p + \chi^*_{-\p} \hat{a}_{-\p}^\dagger, g^{\k_1, \k_2}\hat{a}_{-\k_2}^\dagger \hat{a}_{\k_1}\right]\right]=\\
&-\chi_\q \chi^*_{-\p} g^{\k_1, \k_2} \delta(\k_1+\p)\delta(\q+\k_2)-\chi^*_{-\q} \chi_\p g^{\k_1, \k_2} \delta(\p+\k_2)\delta(\k_1+\q),
\end{split}
\end{equation}
for the third term
\begin{equation}
\begin{split}
&\left[\chi_\q \hat{a}_\q + \chi^*_{-\q} \hat{a}_{-\q}^\dagger, \left[\chi_\p \hat{a}_\p + \chi^*_{\p} \hat{a}_{-\p}^\dagger, g^{*\k_1, \k_2}\hat{a}_{-\k_1}^\dagger \hat{a}_{\k_2}\right]\right]=\\
&-\chi_\q \chi^*_{-\p} g^{*\k_1, \k_2} \delta(\k_2+\p)\delta(\q+\k_1)-\chi^*_{-\q} \chi_\p g^{*\k_1, \k_2} \delta(\p+\k_1)\delta(\k_2+\q),
\end{split}
\end{equation}
and, for the last term 
\begin{equation}
\left[\chi_\q \hat{a}_\q + \chi^*_{-\q} \hat{a}_{-\q}^\dagger, \left[\chi_\p \hat{a}_\p + \chi^*_{-\p} \hat{a}_{-\p}^\dagger, f^{*\k_1, \k_2}\hat{a}_{-\k_1}^\dagger \hat{a}_{-\k_2}^\dagger \right]\right]= \chi_\q \chi_\p f^{*\k_1, \k_2} \left( \delta(\p+\k_2)\delta(\q+\k_1) + \delta(\p+\k_1) \delta(\q+\k_2) \right).
\end{equation}
Collecting all the terms, and integrating over $\k_1, \k_2$, $\p$, $\x$ and $\y$, the correction ${\delta \overline{\mathcal{R}^2}(\eta)}$ is given by
\begin{align}
{\delta \overline{\mathcal{R}^2}(\eta)}=-\frac{\lambda \rC^3}{2m_0^2 \pi^{3/2} \epsilon_{\text{inf}} M_{\mathrm{P}}^2 a^2(\eta)} \int_{\eta_0}^\eta \frac{\dd \eta'}{a^4 (\eta')} \int \dd \q e^{-\frac{\rC^2}{a^2(\eta')}\q^2} &\left[\chi^*_\q (\eta') \chi^*_{-\q} (\eta') f^{-\q,\q}_{\eta}+\chi_{\q}(\eta') \chi_{-\q}(\eta')f^{*\q,-\q}_{\eta}\right.\nonumber\\&\left. -\chi_{\q}(\eta') \chi^*_{\q}(\eta') (g^{\q, -\q}_{\eta}+g^{*-\q,\q}_{\eta})\right].
\end{align}
Using the properties  $f^{-\q,\q}_{\eta}=f^{q,q}_{\eta}$, $g^{-\q,\q}_{\eta}=g^{q,q}_{\eta}$, and choosing the collapse operator as the one taken in Ref.~\cite{Martin2020}, with 
\begin{align}
\chi_{-\k}(\eta')&=\chi_{\k}(\eta')=\chi_{k}(\eta')=\alpha_{k}(\eta')v_k(\eta')+\beta_{k}(\eta')\dot{v}_k(\eta')\label{Chi}\,,\\
\alpha_{k}(\eta) &=\frac{M_{\mathrm{P}}\eta  H^3 \epsilon_{\text{inf}}}{\sqrt{2\epsilon_{\text{inf}}}} \left(-\frac{6 (\epsilon_{\text{inf}} (\epsilon_2/2+1))}{\eta ^2 k^2}+\epsilon_2+8\right)\label{alpha},\\
\beta_k(\eta)&=-\frac{M_{\mathrm{P}}\eta ^2 H^3 \epsilon_{\text{inf}}}{\sqrt{2 \epsilon_{\text{inf}}}}\left(\frac{6 \epsilon_{\text{inf}}}{\eta ^2 k^2}-2\right)\label{beta},
\end{align}
the correction to the mean squared value of the comoving curvature perturbation becomes 
\begin{equation}\label{CurlyFIntLin}
{\delta \overline{\mathcal{R}^2}(\eta_{e})}=-\frac{\lambda \rC^3}{2 m_0^2 \pi^{3/2} \epsilon_{\text{inf}} M_{\mathrm{P}}^2 a^2(\eta_{e})} \int_{\eta_0}^{\eta_{e}} \frac{\dd \eta'}{a^4 (\eta')} \int \dd \q e^{-\frac{\rC^2}{a^2(\eta')}\q^2}\mathcal{F}^q_{\eta'}\,,
\end{equation}
with
\begin{equation}
\mathcal{F}^q_{\eta'}= \chi^*_q (\eta') \chi^*_{q} (\eta') f^{q,q}_{\eta_{e}}+\chi_{q}(\eta') \chi_{q}(\eta')f^{*q,q}_{\eta_{e}}-\chi_{q}(\eta' ) \chi^*_{q}(\eta' ) (g^{q, q}_{\eta_{e}}+g^{*q,q}_{\eta_{e}}).
\end{equation}
During the inflationary era $a^4(\eta)=1/(H^4\eta^4)$, and to leading order in $q\eta'$, $\mathcal{F}^q_{\eta'}$ becomes
\begin{equation}\label{CurlyFLinear}
\mathcal{F}^q_{\eta'}\approx -18 \frac{\epsilon_{\text{inf}}^3 H^6 \eta'^2 M^2_{\mathrm{P}}}{q^{4}\eta_e^2}.
\end{equation}
We notice that for the GRW value of $\rC=1.24\times 10^{27}M^{-1}_{\mathrm{P}}$, the wavelength of the CMB modes $10^{-60}M_{{\mathrm{P}}}\lesssim q\lesssim 10^{-56}M_{{\mathrm{P}}}$ is stretched out on scales greater than $\rC$ at the end of inflation $\eta_e=-10^{34}M^{-1}_{\mathrm{P}}$, leading to $\rC\ll a(\eta_e)/q$. Thus, using Eq.~\eqref{CurlyFLinear} for completing the integral in Eq.~\eqref{CurlyFIntLin}, to leading order we get 
\begin{equation}
{\delta \overline{\mathcal{R}^2}(\eta_{e})}\approx \frac{135}{4}\int \dd \ln q\, \frac{\epsilon_{\text{inf}}^2\lambda H^{5}}{m^2_{0}q^8 \rC^4}.
\end{equation}
The correction to the power spectrum for this choice of linearised collapse operator can be easily read off to be
\begin{equation}
\delta\mathcal{P}_{\mathcal{R}}\approx \frac{\epsilon_{\text{inf}}^2\lambda H^{5}}{m^2_{0}q^8 \rC^4}.
\end{equation} 
We see that for the CMB modes $10^{-60}M_{{\mathrm{P}}}\lesssim q\lesssim 10^{-56}M_{{\mathrm{P}}}$, the correction is hundreds of orders of magnitude larger than the observed value, and is also strongly scale dependent. Our aim in this section is simply to highlight the differences that appear when choosing different collapse operators. When one takes the collapse operator to be quadratic in the perturbations (and henceforth in the creation and annihilation operators), which we took to be proportional to the Hamiltonian density of the perturbations, the dynamical collapse induced correction is negligible. 
However, the linearized collapse operator proportional to the linearized matter-energy density operator, as studied in Ref.~\cite{Martin2020}, indeed leads to much larger corrections inconsistent with observations.

One possible explanation of the different results obtained for the two collapse operators is that the numerical value of the Hamiltonian density of the perturbations $\hat{\mathcal{H}}_{0}(\eta,\x)$ is several orders of magnitude smaller than the matter-energy density $\hat{\delta\rho}$, which can be obtained from $\hat{l}$ in Eq.~\eqref{LinColOp}, by setting $\chi_{\k}(\eta)$ as defined in Eq.~\eqref{Chi} - Eq.~\eqref{beta}. To get a crude estimate of this huge difference during inflation, one can compare $\bra{0}\hat{\mathcal{H}}_{0}(\eta,\x)\ket{0}\sim \bra{0}:\hat{\mathcal{H}}_{0}(\eta,\x):^2\ket{0}^{1/2}\sim \mathcal{O}\left(k^3 |b_\eta^{\k,\k}|\right)$ with $\bra{0}\hat{\delta\rho}^2(\eta,\x)\ket{0}^{1/2}\sim \mathcal{O}\left(k^{3/2} |\chi_{\k}|\right)$   at the time of horizon crossing $\eta= 1/k$, for modes of cosmological interest $k\approx 10^{-60} M_{\mathrm{P}}$. This difference at the level of perturbations is not surprising, since the total energy of the system and the total matter-energy differ by several orders of magnitude, already at the classical level. One can see this by noting that $\bar{\rho}\approx V(\bar{\phi})$ is approximately constant during inflation, and therefore that total matter-energy content increases by a factor of $a^3$. In contrast, the total energy of the system remains conserved, with the increase in matter-energy compensated  by an equal increase in the negative gravitational potential \cite{Guth:2004tw}.

\section{Localization of the wavefunction}

We analyze the claim made in Ref.~\cite{Martin2021}, where it is argued that the power spectrum vanishes for our choice of the collapse operator, namely, the Hamiltonian density of scalar cosmological perturbations. Let us consider a quantum system evolving under a stochastic dynamics controlled by a noise $\xi_t$ with probability distribution $P[\xi]$.  In particular, the standard CSL model \cite{Bassi2003} and the dynamical collapse model considered in our work are two particular cases of this kind of stochastic dynamics. For an operator $\hat{u}$, the expectation value is defined through the relation
\begin{equation}
\bar{u}=\mathbb{E}_\xi \ave{\hat{u}}_\xi = \int \dd \xi P[\xi] \ave{\hat{u}}_\xi.
\end{equation}
In the above expression, $\ave{\hat{u}}_\xi$ denotes the quantum expectation value for a given realization of the noise, and $\mathbb{E}_\xi$ denotes the average with respect to the noise. Let us consider the particular case in which $\bar{u}=0$. In this case, the variance $\sigma_u^2$ for $\hat{u}$ reduces to
\begin{equation}
\sigma_u^2 = \mathbb{E}_\xi \ave{\hat{u}^2}_\xi = \int \dd \xi P[\xi] \ave{ \hat{u}^2 }_\xi.
\end{equation}
Let us assume that for each realisation of the noise, the amplitude square of the wavefunction $\Psi[u]_\xi$ reduces to a Dirac delta centered around some value (which will be equal to the quantum expectation value for that realisation), i.e., {$|\Psi[u]_\xi|^2=\delta(u-\ave{u_{\xi}})$}. In this case, it trivially follows that $\ave{\hat{u}^2}
_\xi=\ave{\hat{u}}_\xi^2$, and moreover
\begin{equation}
{\mathop{\mathbb{E}}}_\xi \langle \hat{u}^2 \rangle_{\xi} \; \stackrel{\left|\Psi[u]_{\xi}\right|^2 = \delta(u - \langle u \rangle_{\xi})}{\longrightarrow} \; {\mathop{\mathbb{E}}}_\xi \langle\hat{u}\rangle_{\xi}^2 \label{locwf}.
\end{equation}

We calculate the power spectrum using the LHS of Eq.~\eqref{locwf}, where no localization assumption is needed. In \cite{Martin2021}, on the contrary, the power spectrum is computed using the RHS of Eq.~\eqref{locwf}, under the assumption of a fully localised wave function. It is clear that the expression used by \cite{Martin2021} is a special case of the standard, more general formula used in our approach. 
Thus, we see that the assumption of Ref.~\cite{Martin2021} does not hold in general, and in fact need not to be applied to calculate the variance, as we have shown in our work. Moreover, our approach is a straightforward generalization to include the effects of the CSL dynamics of the power spectrum definition in standard cosmology [cf. Eq.~\eqref{PowerSpectrumdef}], as the difference is the incorporation of the stochastic average over the realizations. 

Therefore, the claim that the power spectrum of $\hat{\mathcal{R}}$ vanishes {would have been valid only under the assumption of a perfect localization of the wavefunction}, which is not the case considered in our work. Our results show that by choosing a collapse operator which scales with the Hamiltonian density of the scalar perturbations, one can obtain well-defined expressions for the power spectrum, consistent with observations.

It is certainly true that with our choice of the collapse operator, there is no perfect localization of the wavefunction during inflation. However, as the CSL model describes a process continuous in time, it may be argued that the localization can occur later in the evolution of the universe, and not necessarily already at the level of perturbations. Therefore, there is still room to solve the measurement problem in Cosmology.

\end{document}